\documentclass[aip,jmp,reprint,onecolumn]{revtex4-1}

\draft % marks overfull lines with a black rule on the right

\usepackage{graphicx}
\usepackage{amsmath}
\usepackage{amssymb}
\usepackage{bm}
\numberwithin{equation}{section}

\allowdisplaybreaks

\DeclareMathAlphabet{\mathsfbfit}{OT1}{lmss}{bx}{sl}

\begin{document}

%%%%%%%%%%%%%%%%%%%%%%%%%%%%%%%%%%%%%%%%%%%%%%%%%%%%%%%%%%%

\title{$\boldsymbol{\mathsf{\mathsfbfit{SL}(3,\mathsf{C})}}$ structure of one-dimensional Schr\"odinger equation }

\author{Toru Miyazawa}

\email{toru.miyazawa@gakushuin.ac.jp}
\affiliation{Department of Physics, Gakushuin University, Tokyo 171-8588, Japan}
\begin{abstract}
We present a new formalism for describing solutions of the one-dimensional stationary Schr\"odinger equation in terms of the Lie group $SL(3,\mathbf{C})$ and its Lie algebra.
In this formalism, we obtain a universal expression for the Green function which can be used in any representation of $SL(3,\mathbf{C})$ and also expressions for various quantities involving products of Green functions.
Specifically, we introduce an infinite-dimensional representation of $SL(3,\mathbf{C})$ that provides a natural description of multiple scattering of waves.
Using this particular representation, we can derive formulas which are useful for the analysis of the Green function.
\end{abstract}

%\pacs{03.65.Nk, 02.30.Hq, 02.50.Ey}
\maketitle

\section{Introduction}
We consider the one-dimensional stationary Schr\"odinger equation\begin{equation}
-\frac{d^2}{dx^2} \psi (x) + V_\textrm{S}(x) \psi(x) = k^2 \psi(x),
\end{equation}
where the potential $V_{\rm S}(x)$ is a real-valued function defined for $-\infty < x < \infty$ and $k$ is a complex number, $\textrm{Im}\,k \geq 0$.
Shifting $V_\textrm{S}(x)$ by a constant if necessary, we set the origin of the energy scale at the ground-state energy. 
It is well known\cite{schroedinger, infeld, darboux} that 
 the Schr\"odinger equation can be written as a first-order equation with two components,
\begin{equation}
\frac{d}{dx}
\begin{pmatrix}
\psi \\
\phi
\end{pmatrix}
=
\begin{pmatrix}
f(x) & -ik \\
-ik  & -f(x) \\
\end{pmatrix}
\begin{pmatrix}
\psi \\
\phi
\end{pmatrix}.
\end{equation}
It follows from (1.2) that
\begin{equation}
\frac{d^2}{dx^2}\psi = (f^2 + f') \psi  - k^2 \psi, \qquad
\frac{d^2}{dx^2}\phi = (f^2 - f') \phi  - k^2 \phi ,
\end{equation}
where $f'$ denotes the derivative of $f$. 
Thus, (1.2) includes two Schr\"odinger equations as a pair.
We will use only the first equation of (1.3), so we identify $V_\mathrm{S}(x)$ as
\begin{equation}
V_{\rm S}(x)=[f(x)]^2 + f'(x).
\end{equation}
Let $\psi_0(x)$ be a function
\cite{Note1}
satisfying (1.1) with $k^2=0$. We can choose $\psi_0$ such that $\psi_0(x) > 0$ for any $x$. The function $f$ is obtained from $\psi_0$ as $f(x) = \psi_0'(x)/\psi_0(x)$. [This can be checked by substituting into (1.4).] 

For later convenience, we rewrite (1.2) as
\begin{equation}
\frac{d}{dx}
\begin{pmatrix}
\psi + \phi \\
\psi - \phi
\end{pmatrix}
=
\begin{pmatrix}
-ik & f(x) \\
f(x)  & ik \\
\end{pmatrix}
\begin{pmatrix}
\psi + \phi \\
\psi - \phi
\end{pmatrix}.
\end{equation}
Let the $2 \times 2$ matrix $U(x,y)$ be defined as the solution of
\begin{equation}
\frac{\partial}{\partial x}
U(x,y)
=
\begin{pmatrix}
-ik & f(x) \\
f(x)  & ik \\
\end{pmatrix}
U(x,y)
\end{equation}
with the initial condition $U(y,y)=I$, where $I$ denotes the $2 \times 2$ unit matrix.
The four elements of the matrix $U(x,y)$ can be written in terms of two functions $\alpha$ and $\beta$ as
\begin{equation}
U(x,y)=
\begin{pmatrix}
\alpha(x,y;k) & \beta(x,y;-k) \\
\beta(x,y;k) & \alpha(x,y;-k)
\end{pmatrix}.
\end{equation}
By comparing (1.6) with (1.5), we can see that the functions $\psi_-$ and $\psi_+$ defined by
\begin{equation}
\psi_-(x,y) \equiv \alpha(x,y;k) + \beta(x,y;k), 
\qquad
\psi_+(x,y) \equiv \alpha(x,y;-k) + \beta(x,y;-k)
\end{equation}
are two solutions of the Schr\"odinger equation (1.1). Any solution $\psi(x)$ of the Schr\"odinger equation can be expressed as a linear combination of $\psi_+(x,y)$ and $\psi_-(x,y)$ with a fixed $y$.

Since the matrix in front of $U$ on the right-hand side of (1.6) is traceless, the determinant of $U(x,y)$ does not depend on $x$. From $U(y,y) = I$, it follows that $\det U(x,y) = 1$.
The matrix $U$ is a $2 \times 2$ matrix with unit determinant, and so we can interpret it as belonging to the Lie group $SL(2,\mathbf{C})$. 
Thus, the Schr\"odinger equation has an obvious $SL(2,\mathbf{C})$ structure. 
The matrix in front of $U$ in (1.6) can be viewed as an element of the Lie algebra $sl(2,\mathbf{C})$, which consists of $2 \times 2$ traceless matrices.
As a basis of $sl(2,\mathbf{C})$, we can choose
\begin{equation}
J_1 = \frac{-1}{2} 
\begin{pmatrix}
\,0 & 1 \,\\
\,1 & 0 \,
\end{pmatrix}, \qquad
J_2 = \frac{1}{2} 
\begin{pmatrix}
\,0 & -i \\
\,i & 0 
\end{pmatrix}, \qquad
J_3 = \frac{1}{2} 
\begin{pmatrix}
-1 & 0 \,\\
0 & 1 \,
\end{pmatrix}.
\end{equation}
These matrices satisfy the commutation relations
\begin{equation}
[J_1, J_2] = i J_3, \qquad [J_2, J_3] = i J_1, \qquad [J_3, J_1] = i J_2.
\end{equation}
Using (1.9), the right-hand side of (1.6) can be written as $(2 i k J_3 - 2 f J_1)\,U$. 
This $SL(2,\mathrm{C})$ structure of the Schr\"odinger equation is well known, and it is used in connection with the factorization of the Schr\"odinger operator and the supersymmetric formulation of quantum mechanics.\cite{witten, cooper}

Needless to say, the Schr\"odinger equation is one of the most extensively studied equations in physics, and there are many approaches to the analysis of its solutions. 
The aim of this paper is to propose a new formalism which enables us to study the one-dimensional Schr\"odinger equation from a totally different point of view.
In this paper, we consider an extension of the $SL(2,\mathrm{C})$ structure by embedding it in $SL(3,\mathrm{C})$. 
This $SL(3,\mathrm{C})$ formalism is useful for describing solutions of the Schr\"odinger equation and, in particular, the Green function.
Within this framework, we can construct a general expression for the Green function which includes various already-known expressions as specific representations. 
For example, in the most commonly used construction of the Green function, it is expressed in terms of Jost solutions or other solutions satisfying the required boundary conditions. Another important description of the Green function is given by its expression as a superposition of multiply scattered waves.
These two apparently different pictures of the Green function are, in fact, merely two different representations of our general expression. 
In this way, our formalism provides a more general viewpoint on the Schr\"odinger equation, with which the structure of its solutions can be more clearly understood than in previous approaches.

The $SL(3,\mathrm{C})$ formalism is not only of theoretical interest. 
By making use of the $SL(3,\mathrm{C})$ structure, we can obtain various formulas which are notably useful for the analysis in low- and high-energy regions. 
Analytic behavior of solutions of the Schr\"odinger equation has been studied over the years by numerous researchers both in the low-energy region\cite{deift, yafaef, bolle, newton, klaus1, aktosun3, costin} and in the high-energy region.\cite{verde, harris1, hinton1, rybkin2, ramond} 
Compared to the conventional methods used in these works, our formalism is especially suited for examining the Green function.
Using the formulas derived in this paper, the low-energy and the high-energy behavior of the Green function can be investigated more systematically than by other methods. In the present paper, however, we confine ourselves to the formal aspect of the theory. 
Its application to the analysis of solutions will be discussed elsewhere. 

Our formulation does not depend on the specific form of the potential, so $V_\mathrm{S}(x)$ in (1.1) and $f(x)$ in (1.2) may be quite general. For simplicity, we assume that $f(x)$ is a piecewise continuously differentiable function of $x$. [The function $f(x)$ may have jump discontinuities, and so $V_\mathrm{S}(x)$ given by (1.4) may have delta-function singularities.] No further conditions of smoothness are required.
As for the behavior at infinity, we assume that the limits $f(\infty) \equiv \lim_{x \to + \infty} f(x)$ and $f'(\infty) \equiv \lim_{x \to + \infty} f'(x)$ exist. These limits may be infinite. That is to say, $f(\infty)$ is finite, $+ \infty$, or $-\infty$ [and similarly for $f'(\infty)]$. 
From this assumption and (1.4), it follows that $V_\mathrm{S}(\infty)$ is either finite or $+\infty$. 
We make the same assumption for the behavior of $f(x)$ as $x \to -\infty$. Namely, $f(-\infty)$ is finite, $+\infty$, or $-\infty$ [and similarly for $f'(-\infty)$]. Accordingly, $V_\mathrm{S}(-\infty)$ is either finite or $+ \infty$. We do not consider the cases where $V_\mathrm{S}(\infty) = -\infty$ or $V_\mathrm{S}(-\infty) = -\infty$.
Our method can be applied to more general cases including, for example, asymptotically periodic potentials and potentials that diverge at finite $x$. However, to avoid complication, we do not discuss such cases in this paper.

\vfill

\section{Transmission and reflection coefficients}

It is obvious that  the matrix $U$ defined in Sec.~I satisfies  $U(x,y) U(y,z)=U(x,z)$. Hence,\begin{align}
\alpha(x,z;\pm k) 
&= \alpha(x,y;\pm k) \alpha(y,z;\pm k) 
+ \beta(x,y;\mp k) \beta(y,z;\pm k),
\nonumber \\
\beta(x,z;\pm k)
&= \beta(x,y;\pm k) \alpha(y,z;\pm k) + \alpha(x,y;\mp k) \beta(y,z;\pm k).
\end{align}
The condition $U(x,x)=I$ means $\alpha(x,x;\pm k)=1$ and $\beta(x,x;\pm k)=0$. 
The determinant of $U$ is unity, so $\alpha$ and $\beta$ satisfy
\begin{equation}
\alpha(x,y;k) \alpha(x,y;-k) -\beta(x,y;k) \beta(x,y;-k)=1.
\end{equation}
Since $U(x,y) U(y,x)=U(x,x) = I$, the matrix $U(y,x)$ is the inverse matrix of $U(x,y)$. 
Equating the elements of $U(y,x)$ and $[U(x,y)]^{-1}$ gives
\begin{equation}
\alpha(y,x;\pm k) = \alpha(x,y; \mp k), \qquad
\beta(y,x;\pm k) = - \beta(x,y; \pm k).
\end{equation}
Using (2.3), the functions $\psi_\pm$ defined by (1.8) can also be written as
\begin{equation}
\psi_+(x,y) = \alpha(y,x;k) - \beta(y,x;-k).
\qquad 
\psi_-(x,y) = \alpha(y,x;-k) - \beta(y,x;k), 
\end{equation}

It can be seen  from (1.6) and (1.7) that $\alpha(x,y; \pm k) = e^{\mp i k (x-y)}$ and $\beta(x,y;\pm k) = 0$ if $f$ is identically zero. 
Therefore, $\psi_\pm (x,y) = e^{\pm ik(x-y)}$ if $f=0$.
Let $x_1$ and $x_2$ be fixed numbers, $x_2 \geq x_1$, and suppose that $f(x)=0$ for $x<x_1$ and $x>x_2$. Then, 
\begin{equation}
\psi_{\pm}(x,x_1)= e^{\pm ik(x-x_1)} \quad \mbox{for} \ \ x<x_1, \qquad
\psi_{\pm}(x,x_2)= e^{\pm ik(x-x_2)} \quad \mbox{for} \ \ x>x_2.
\end{equation} 
From (2.1), (1.8), and (2.3), we can easily derive the relations
\begin{subequations}
\begin{align}
\psi_+(x,x_2) &=\alpha(x_2,x_1;k) \psi_+(x,x_1) - \beta(x_2,x_1;-k)\psi_-(x,x_1),
\\
\psi_-(x,x_1) &=\alpha(x_2,x_1;k) \psi_-(x,x_2) + \beta(x_2,x_1;k)\psi_+(x,x_2).
\end{align}
\end{subequations}
Let us define the transmission coefficient $\tau$, the right reflection coefficient $R_r$, and the left reflection coefficient $R_l$ for the interval $[x_1,x_2]$ as
\begin{equation}
\tau(x_2,x_1) \equiv \frac{1}{\alpha(x_2,x_1;k)}, \qquad
R_r(x_2,x_1) \equiv  \frac{\beta(x_2,x_1;k)}{\alpha(x_2,x_1;k)}, \qquad
R_l(x_2,x_1) \equiv - \frac{\beta(x_2,x_1;-k)}{\alpha(x_2,x_1;k)}.
\end{equation}
(We omit to write the dependence of $\tau$, $R_r$, and $R_l$ on $k$.)
We can rewrite (2.6) as
\begin{subequations}
\begin{align}
\phi_1(x) \equiv \tau(x_2,x_1) \psi_+(x,x_2) &= \psi_+(x,x_1)  + R_l(x_2,x_1)\psi_-(x,x_1),
\\
\phi_2(x) \equiv \tau(x_2,x_1) \psi_-(x,x_1) &= \psi_-(x,x_2) + R_r(x_2,x_1)\psi_+(x,x_2).
\end{align}
\end{subequations}
We find from (2.8) and (2.5) that 
\begin{subequations}
\begin{align}
\phi_1(x) 
&= 
\Biggl\{
\begin{array}{ll}
\tau(x_2,x_1) e^{ik(x-x_2)}, & \quad x>x_2, \\[1ex]
e^{ik(x-x_1)} + R_l(x_2,x_1) e^{-ik(x-x_1)}, & \quad x<x_1,
\end{array}
\\
\phi_2(x) 
&=
\Biggl\{
\begin{array}{ll}
e^{-ik(x-x_2)} + R_r(x_2,x_1) e^{ik(x-x_2)}, & \quad x>x_2, \\[1ex]
\tau(x_2,x_1) e^{-ik(x-x_1)}, & \quad x<x_1.
\end{array}
\end{align}
\end{subequations}
%
%
%       Figure 1
%
\begin{figure}
\includegraphics[scale=0.7]{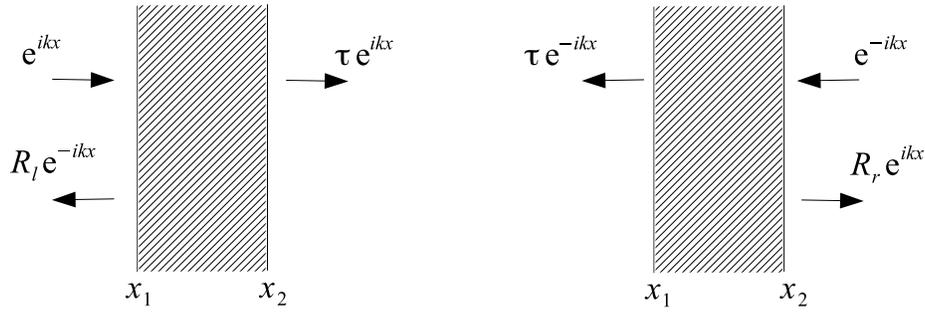}
\caption{
Schematic illustration of (2.9). 
(For simplicity, the factors $e^{\pm i k x_1}$ and $e^{\pm i k x_2}$ are omitted.)
}
\end{figure}
Hence we can see that $\tau$, $R_r$, and $R_l$ are indeed the transmission and reflection coefficients for the interval $[x_1,x_2]$, as illustrated in Fig.~1. 

We define the transmission and reflection coefficients for an interval $[x_1,x_2]$ by (2.7) for general $f(x)$ which is not necessarily zero outside this interval. We define the truncated function
\begin{equation}
\bar f(x) \equiv \chi_{[x_1,x_2]}(x)f(x),
\qquad
\chi_{[x_1,x_2]}(x) \equiv
\Biggl\{
\begin{array}{ll}
1, & \ x_1 \leq x \leq x_2,\\[1ex]
0, & \ \text{otherwise}
\end{array}
\end{equation}
and consider the corresponding  potential [see Eq.~(1.4)] for the Schr\"odinger equation
\begin{equation}
\bar V_{\rm S}(x) \equiv [\bar f(x)]^2 + (d/dx)\bar f(x) = \chi_{[x_1,x_2]}(x)\,V_{\rm S}(x) - f(x_2)\, \delta(x-x_2) + f(x_1)\, \delta(x-x_1).
\end{equation}
Obviously, $\bar V_\mathrm{S}(x)=V_\mathrm{S}(x)$ for $x_1<x<x_2$, and $\bar V_\mathrm{S}(x)=0$ for $x<x_1$ and $x_2<x$. In addition, $\bar V_{\rm S}(x)$ includes delta functions at $x=x_1$ and $x=x_2$ unless $f(x)$ happens to vanish there.
The quantities $\tau(x_2,x_1)$, $R_r(x_2,x_1)$, and $R_l(x_2,x_1)$ defined by (2.7) can be thought of as the transmission and reflection coefficients for the Schr\"odinger equation with this potential~$\bar V_{\rm S}$. 

The transmission and reflection coefficients can be defined for any closed interval, and they are functions of the two endpoints of the interval.
From $U(x,x)=I$, we have $\tau(x,x) = 1$ and $R_r(x,x) = R_l(x,x) = 0$.
We can also define
\cite{Note2}
the reflection coefficients for semi-infinite intervals by 
$R_r(x,-\infty) \equiv \lim_{z \to -\infty}  R_r(x,z)$ and 
$R_l(\infty,x) \equiv \lim_{z \to \infty} R_l(z,x)$.

\section{Algebraic construction of solutions}
As already noted, the right-hand side of (1.6) can be written as
$(2ik J_3 - 2 f J_1) U$ with (1.9). 
Let us consider a more general case. We assume that $J_1$, $J_2$, and $J_3$ are some linear operators (or matrices) satisfying the commutation relations (1.10). 
They are not necessarily the $2 \times 2$ matrices of (1.9). 
For such $J_1$ and $J_3$, we define the evolution operator $U(x,y)$ as the solution of  
\begin{equation}
\frac{\partial}{\partial x} U(x,y) = [2ik J_3 - 2 f(x) J_1] U(x,y)
\end{equation}
with the initial condition
\begin{equation}
U(y,y)=1,
\end{equation}
where $1$ denotes the identity operator.
Equation (3.1) is a generalization of (1.6), and $U(x,y)$ is an operator that acts on some vector space. 
Here and hereafter, $U(x,y)$ is not necessarily the $2 \times 2$ matrix of Secs.~I and II. 
The following relations, which we used in Sec.~II for $2 \times 2$ matrices, hold for the general case as well:
\begin{equation}
U(x,y)U(y,z)=U(x,z), \qquad U(y,x)=[U(x,y)]^{-1}.
\end{equation}
We define the raising and lowering operators $J_\pm$ by
\begin{equation}
J_+ \equiv J_1 + i J_2, \qquad
J_- \equiv J_1 - i J_2.
\end{equation}
The $sl(2, \mathbf{C})$ commutation relations (1.10) then read
\begin{equation}
[J_+, J_-] = 2 J_3, \qquad [J_3, J_+] = J_+, \qquad [J_3, J_-] = - J_-.
\end{equation}
The operator $U$ satisfying (3.1) and (3.2) can be expressed using $J_\pm$ and $J_3$ as
\begin{equation}
U(x,y) = \exp[-R_r(x,y) J_+] [\tau(x,y)]^{2 J_3} \exp[R_l(x,y) J_-],
\end{equation}
where
$\tau^{2 J_3} = \exp[2 (\log \tau) J_3]$. 
The proof of (3.6) is given in Appendix~A.

Suppose that there exist operators $Q_+$ and $Q_-$ that satisfy the commutation relations
\begin{alignat}{4}
[Q_+, J_3] &= -\frac{1}{2} Q_+, &\qquad 
[Q_+, J_+] &= 0, &\qquad 
[Q_+, J_-] &= - Q_-,
\nonumber 
\\
[Q_-, J_3] &= \frac{1}{2} Q_-, &\qquad
[Q_-, J_+] &= - Q_+, &\qquad
[Q_-, J_-] &= 0.
\end{alignat}
Examples of such operators will be shown later.
We can also write~(3.7) as
\begin{gather}
\begin{pmatrix}
[Q_+, J_3] \\
[Q_-, J_3]
\end{pmatrix}
= \frac{1}{2}
\begin{pmatrix}
-1 & 0 \,\\
0 & 1 \,
\end{pmatrix}
\begin{pmatrix}
Q_+ \\
Q_-
\end{pmatrix},
\nonumber
\\[1.5ex]
\begin{pmatrix}
[Q_+, J_+] \\
[Q_-, J_+]
\end{pmatrix}
= 
\begin{pmatrix}
0 & 0 \,\\
-1 & 0 \,
\end{pmatrix}
\begin{pmatrix}
Q_+ \\
Q_-
\end{pmatrix}, \qquad
\begin{pmatrix}
[Q_+, J_-] \\
[Q_-, J_-]
\end{pmatrix}
= 
\begin{pmatrix}
\,0 & -1 \\
\,0 & 0
\end{pmatrix}
\begin{pmatrix}
Q_+ \\
Q_-
\end{pmatrix}.
\end{gather}
Note that the $2 \times 2$ matrices appearing on the right-hand sides of (3.8) are the matrices for $J_3$ and $J_1 \pm i J_2$ given by (1.9).
From (3.7), it follows that
\begin{alignat}{4}
Q_+ e^{c J_+} &= e^{c J_+} Q_+, &\qquad 
Q_+ c^{2J_3} &= c^{2 J_3 - 1} Q_+, &\qquad
Q_+ e^{c J_-} &= e^{c J_-}(Q_+ - c Q_-), 
\nonumber \\
e^{c J_-} Q_- &=  Q_- e^{c J_-}, &\qquad
c^{2 J_3} Q_- &= Q_-c^{2 J_3 - 1}, &\qquad
e^{c J_+} Q_- &= (Q_- + c Q_+) e^{c J_+},
\end{alignat}
where $c$ is an arbitrary complex number.
Using (3.9) successively with (3.6) yields
\begin{subequations}
\begin{align}
Q_+ U 
&= e^{-R_r J_+} Q_+ \tau^{2 J_3} e^{R_l J_-}
= e^{-R_r J_+} \tau^{2 J_3 - 1} Q_+ e^{R_l J_-}
=(1/\tau) U (Q_+ - R_l Q_-),
\\
U Q_- &= e^{-R_r J_+} \tau^{2 J_3} Q_- e^{R_l J_-} 
= e^{- R_r J_+} Q_- \tau^{2 J_3 - 1} e^{R_l J_-}
= (1/\tau) (Q_- - R_r Q_+) U.
\end{align}
\end{subequations}
Multiplying Eqs.~(3.10) by $\tau$, and rearranging the terms, we obtain
\begin{subequations}
\begin{align}
&U(x_2,x_1) Q_+ = \tau(x_2,x_1) Q_+ U(x_2,x_1) + R_l(x_2,x_1) U(x_2,x_1) Q_-,
\\
&Q_- U(x_2,x_1) = \tau(x_2,x_1) U(x_2,x_1) Q_- + R_r(x_2,x_1) Q_+ U(x_2,x_1).
\end{align}
\end{subequations}
The structure of Eqs.~(3.11) is illustrated in Fig.~2.
%
%
%       Figure 2
%
\begin{figure}
\includegraphics[scale=0.7]{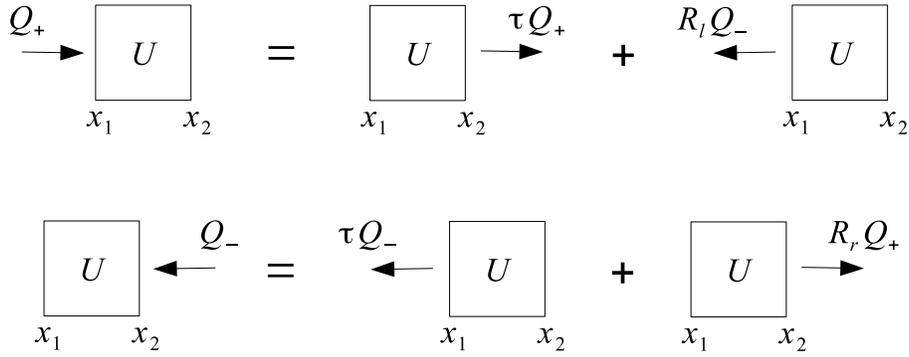}
\caption{
Schematic illustration of Eqs.~(3.11). 
(The left-right direction is reversed in the equations.)
}
\end{figure}
Comparing Fig.~2 with Fig.~1, we find that Eqs.~(3.11) can be viewed as a mathematical representation of the scattering events shown in Fig.~1.
In each equation of (3.11), the left-hand side describes the incident wave, while the right-hand side describes the transmitted and reflected waves.
The operators $Q_+$ and $Q_-$ correspond to the right-going wave $e^{i k x}$ and the left-going wave $e^{-i k x}$, respectively. 
Substituting (2.7), we can rewrite (3.11) in another form,
\begin{subequations}
\begin{align}
&Q_+ U(x_2,x_1) = U(x_2,x_1) [ \alpha(x_2,x_1;k) Q_+ + \beta(x_2,x_1;-k)Q_-], 
\\
&U(x_2,x_1) Q_- = [\alpha(x_2,x_1;k) Q_- - \beta(x_2,x_1;k) Q_+] U(x_2,x_1).
\end{align}
\end{subequations}
Interchanging $Q_+$ and $Q_-$, and changing the sign of $k$ in $\alpha$ and $\beta$, we obtain\begin{subequations}
\begin{align}
&Q_- U(x_2,x_1) = U(x_2,x_1) [ \alpha(x_2,x_1;-k) Q_- + \beta(x_2,x_1;k)Q_+],
\\
&U(x_2,x_1) Q_+ = [\alpha(x_2,x_1;-k) Q_+ - \beta(x_2,x_1;-k) Q_-] U(x_2,x_1).
\end{align}
\end{subequations}
Equations~(3.13) can also be derived from (3.12) by direct calculation using (2.2).

From (3.1) and $(\partial/\partial x) [U(z,x) U(x,y)] = (\partial/\partial x) U(z,y) =0$, it follows that
\begin{equation}
\frac{\partial}{\partial x}U(z,x)=U(z,x)[-2ik J_3 + 2 f(x) J_1].
\end{equation}
Let $A$ be an arbitrary operator. 
From (3.1) and (3.14) we have
\begin{equation}
\frac{\partial}{\partial x}U(z,x) A U(x,z') =2ik U(z,x)[A, J_3]U(x,z') - 2 f(x) U(z,x) [A, J_1]U(x,z'),
\end{equation}
where $z$ and $z'$ are arbitrary real numbers.
Setting $A=Q_\pm$, and using (3.8), we obtain
\begin{equation}
\frac{\partial}{\partial x}
\begin{pmatrix}
U(z,x) Q_+ U(x,z') \\
U(z,x) Q_- U(x,z')
\end{pmatrix}
=
\begin{pmatrix}
-ik & f(x) \\
f(x) & ik
\end{pmatrix}
\begin{pmatrix}
U(z,x) Q_+ U(x,z') \\
U(z,x) Q_- U(x,z')
\end{pmatrix}.
\end{equation}
This equation has the same form as (1.5). 
We can also rewrite (3.16) in the form of (1.2),
\begin{equation}
\frac{\partial}{\partial x}
\begin{pmatrix}
U(z,x) (Q_+ + Q_-) U(x,z') \\
U(z,x) (Q_+ - Q_-) U(x,z')
\end{pmatrix}
=
\begin{pmatrix}
f(x) & -ik \\
-ik & -f(x)
\end{pmatrix}
\begin{pmatrix}
U(z,x) (Q_+ + Q_-) U(x,z') \\
U(z,x) (Q_+ - Q_-) U(x,z')
\end{pmatrix}.
\end{equation}

It is our assumption that $U$ and $Q_\pm$ are operators (or matrices) acting on some vector space. 
We assume that this vector space is equipped with an inner product.
Let $\langle \Psi_2, \Psi_1 \rangle$ denote the inner product of two vectors $\Psi_1$ and $\Psi_2$. This inner product is defined to be linear in the right entry and conjugate linear in the left entry. The adjoint of an operator $A$, denoted by $A^\dagger$, is defined by 
$\langle \Psi_2, A \Psi_1 \rangle = \langle A^\dagger \Psi_2, \Psi_1 \rangle$.

Let $\Psi$ and $\Phi$ be some fixed vectors in this vector space. We define
\begin{equation}
\psi(x;z,z') \equiv 
\langle \Phi, U(z,x) (Q_+ + Q_-) U(x,z') \Psi \rangle.
\end{equation}
As can be seen from (3.17), this $\psi$ satisfies the same equation as the $\psi$ in (1.2).  Therefore, this $\psi(x;z,z')$, as a function of $x$, is a solution of the Schr\"odinger equation (1.1). 
It can be expressed as a linear combination of the functions $\psi_\pm$ defined by (1.8). Using (3.12b) and (3.13b) in (3.18), and also using (2.4) and $U(z,x)U(x,z')=U(z,z')$, we have
\begin{equation}
\psi(x; z,z')= \langle \Phi, Q_- U(z,z') \Psi \rangle \psi_+(x,z)
+ \langle \Phi, Q_+ U(z,z') \Psi \rangle \psi_-(x,z).
\end{equation}
Similarly, using (3.12a), (3.13a), and (1.8) gives
\begin{equation}
\psi(x;z,z')=\langle \Phi,  U(z,z') Q_- \Psi \rangle \psi_+(x,z') 
+ \langle \Phi,  U(z,z') Q_+ \Psi \rangle \psi_-(x,z'). 
\end{equation}
The vectors $\Psi$ and $\Phi$ determine the boundary conditions at $x=z'$ and $x=z$, respectively. For example,
$\psi(z';z,z')=0$ if $(Q_+ + Q_- )\Psi = 0$, and $\psi(z; z,z')=0$ if $(Q_+ + Q_-)^\dagger\Phi = 0$.
In this way, boundary conditions can be translated into conditions on $\Psi$ and $\Phi$. 

We can impose boundary conditions at infinity by letting $z \to +\infty$ or $z' \to -\infty$. 
Before thinking about the conditions on $\Psi$ and $\Phi$, we need to study the behavior of $\psi_\pm$ in these limits. 
Let us first assume that $f(x)$ tends to zero sufficiently rapidly as $x \to \pm \infty$. 
Then, the functions $\psi_\pm(x,z)$ behave like $e^{\pm ik(x-z)}$ when both $x$ and $z$ are large. Similarly, $\psi_\pm (x,z')$ behave like $e^{\pm ik(x-z')}$ when both $-x$ and $-z'$ are large. [Recall that $\psi_\pm(x,y) = e^{\pm k (x - y)}$ if $f$ is identically zero.]
If we define
\begin{equation}
\chi_+(x) \equiv \lim_{z \to +\infty} e^{i k z} \psi_+(x,z), \qquad
\chi_-(x) \equiv \lim_{z' \to -\infty} e^{-i k z'} \psi_-(x,z'),
\end{equation}
then $\chi_+$ and $\chi_-$ are the solutions of the Schr\"odinger equation such that
\begin{equation}
\chi_+(x) \sim e^{i k x} \quad \mbox{as} \ x \to +\infty, 
\qquad
\chi_-(x) \sim e^{-i k x} \quad \mbox{as} \ x \to -\infty. 
\end{equation}
The limit $z \to +\infty$ of $\psi_+(x,z)$ does not exist as a finite number. 
In order to obtain a definite function of $x$ in this limit, it is necessary to multiply $\psi_+(x,z)$ by the $z$-dependent normalization factor $e^{i k z}$ before letting $z \to +\infty$, as in (3.21).
Let $x_0$ be a fixed number.
It can be shown
\cite{Note3}  
that $\tau(z,x_0) \sim C_1\, e^{ikz}$ as $z \to +\infty$, where $C_1$ is a $z$-independent factor. 
[Similarly, $\tau(x_0, z') \sim C_2\, e^{-ikz'}$ as $z' \to -\infty$.]
Therefore, we can use $\tau(z,x_0)$ as the normalization factor for $\psi_+(x,z)$ in place of $e^{i k z}$. That is, instead of (3.21) we may define
\begin{equation}
\phi_+(x) \equiv \lim_{z \to +\infty} \tau(z,x_0) \psi_+(x,z), \qquad
\phi_-(x) \equiv \lim_{z' \to -\infty} \tau(x_0,z') \psi_-(x,z').
\end{equation}
The fixed number $x_0$ in (3.23) is arbitrary. By changing the choice of $x_0$, the form of $\phi_\pm$ changes only by a constant factor. It is obvious that $\phi_\pm$ satisfy the boundary conditions
\begin{equation}
\phi_+(x) \sim C_1\, e^{i k x} \quad \mbox{as $x \to +\infty$}, 
\qquad
\phi_-(x) \sim C_2\, e^{-i k x} \quad \mbox{as $x \to -\infty$}, 
\end{equation}
where $C_1$ and $C_2$ are constants that depend on the choice of $x_0$.

Now let us consider the general case where $f(x)$ does not necessarily vanish as $x \to \pm \infty$. Whereas (3.21) is meaningful only when $f(\pm \infty) = 0$, definition (3.23) is valid even if $f(\pm \infty) \neq 0$. 
It is shown in Appendix~B that the limits on the right-hand sides of (3.23) exist as functions of $x$, and that these functions satisfy the boundary conditions
\begin{equation}
\lim_{x \to +\infty} \phi_+(x) = 0, \qquad
\lim_{x \to -\infty} \phi_-(x) = 0 \qquad \mbox{for} \ \  \textrm{Im}\,k >0.
\end{equation}
For real $k$, we may regard $\phi_\pm$ as $\phi_\pm(x;k) = \lim_{\epsilon \downarrow 0}\phi_\pm(x;k + i \epsilon)$. 
As a solution of the Schr\"odinger equation, $\phi_+(x;k)$ is uniquely determined (apart from a constant factor) by the condition $\phi_+(+\infty;k + i \epsilon) = 0$ with $\epsilon > 0$.
Similarly, $\phi_-(x;k)$ is determined by the condition $\phi_-(-\infty;k + i \epsilon) = 0$. 
Thus, (3.25) is sufficient for specifying the boundary conditions even for $\mathrm{Im}\,k = 0$.
In the case $f(\pm \infty) = 0$ discussed above, (3.25) is equivalent to (3.24). 
If $\vert f(\pm \infty) \vert = \infty$, it follows from (3.25) that $\phi_\pm(\pm \infty) = 0$ for $\mathrm{Im}\,k = 0$ as well.

It can be seen from (3.19) that $\psi(x;z,z') \propto \psi_+(x,z)$ if $Q_+^\dagger \Phi = 0$, and from (3.20) that $\psi(x;z,z') \propto \psi_-(x,z')$ if $Q_- \Psi = 0$. 
We find
\begin{subequations}
\begin{align}
\lim_{z \to +\infty} & N_1(z,z') \psi(x;z,z')=\phi_+(x) 
\qquad
\text{if} \ \  Q_+^\dagger \Phi = 0,
\\
\lim_{z' \to -\infty} & N_2(z,z') \psi(x;z,z') = \phi_-(x) 
\qquad
\text{if} \ \  Q_- \Psi = 0,
\end{align}
\end{subequations}
where $N_1 = \tau(z,x_0)/\langle \Phi, Q_- U(z,z') \Psi \rangle$, 
$N_2 = \tau(z_0,z')/\langle \Phi,  U(z,z') Q_+ \Psi \rangle$.
In this way, the solutions of the Schr\"odinger equation satisfying the boundary conditions (3.25) can be obtained from $\psi(x;z,z')$ by imposing the condition $Q_+^\dagger \Phi = 0$ or $Q_- \Psi = 0$, and then taking the limit $z \to + \infty$ or $z' \to -\infty$ with an appropriate normalization factor depending on $z$ and $z'$.
[The conditions $Q_+^\dagger \Phi = 0$ and $Q_- \Psi = 0$ in (3.26) are sufficient conditions. In many cases, (3.26a) and (3.26b) hold under less stringent conditions on $\Psi$ and $\Phi$. See Appendix~D.]

\section{Green function}

We define the Green function $G(x,y;k)$ of the Schr\"odinger equation as the function satisfying
\begin{equation}
\left[- \frac{\partial^2}{\partial x^2} + V_\textrm{S}(x) - k^2 \right] G(x,y;k) = -\delta(x-y)
\end{equation}
with the boundary conditions $G(x,y;k) \to 0$ as $\vert x-y \vert \to \infty$ for $\textrm{Im}\, k>0$. The Green function for real $k$ is defined by $G(x,y;k) \equiv \lim_{\epsilon \downarrow 0} G(x,y;k + i \epsilon)$. 

Let $\Psi_0$ and $\Phi_0$ be vectors satisfying $Q_- \Psi_0 = 0$ and $Q_+^\dagger \Phi_0 = 0$, let $\Psi$ and $\Phi$ be arbitrary vectors, and let $a$ and $a'$ be fixed numbers.
Equations~(3.26) mean that
\begin{subequations}
\begin{alignat}{3}
&\langle \Phi_0, U(z,x) (Q_+ + Q_-) U(x,a') \Psi \rangle
\sim A_1(z) \phi_+(x)
&\qquad &\text{as} \ \  z \to +\infty, \\
&\langle \Phi, U(a,x) (Q_+ + Q_-) U(x,z') \Psi_0 \rangle
\sim A_2(z') \phi_-(x)
&\qquad &\text{as} \ \  z' \to -\infty,
\end{alignat}
\end{subequations}
where $A_1(z)$ and $A_2(z')$ are some functions which also depend on the choice of $\Psi$,  $\Phi$, $a$, and $a'$.
Let $x \geq y$, and consider
\begin{equation}
g_0(x,y;z,z')  \equiv \langle \Phi_0, U(z,x) (Q_+ + Q_-) U(x,y) (Q_+ + Q_-) U(y,z') \Psi_0 \rangle.
\end{equation}
This $g_0$ has the form of (4.2a) as a function of $x$ and (4.2b) as a function of $y$.
Therefore, 
\begin{equation}
\lim_{\mathstrut z \to +\infty}
\lim_{\mathstrut z' \to -\infty}N(z,z') g_0(x,y;z,z')= C \phi_+(x) \phi_-(y),
\end{equation}
where $N(z,z')$ is some appropriate factor, and $C$ is a finite constant. 
This $N(z,z')$ is to be determined later.
With such a factor $N(z,z')$, we define
\begin{equation}
g(x,y;z,z') 
\equiv
\Biggl\{
\begin{array}{ll}
N(z,z') g_0(x,y;z,z'), 
& \quad x \geq y,
\\[1ex]
N(z,z') g_0(y,x;z,z'),
& \quad y \geq x.
\end{array}
\end{equation}
The Green function $G(x,y;k)$ (hereafter we will omit the argument $k$)  is then obtained as
\begin{equation}
G(x,y) = \lim_{\mathstrut z \to + \infty} \lim_{\mathstrut z' \to -\infty} g(x,y;z,z'),
\end{equation}
which reads, with (4.4) and (4.5),
\begin{equation}
G(x,y) =
\Biggl\{
\begin{array}{ll}
C \phi_+(x) \phi_-(y),
& \quad x \geq y,
\\[1ex]
C \phi_+(y) \phi_-(x),
& \quad  y \geq x.
\end{array}
\end{equation}
This $G(x,y)$ indeed satisfies (4.1) for $x \neq y$ and also satisfies the required boundary conditions at infinity [see (3.25)]. 
It only remains to determine the factor $N(z,z')$ so that the delta function on the right-hand side of (4.1) may be correctly produced. 
This requires that\begin{equation}
\lim_{\epsilon \downarrow 0}
\left[
\frac{\partial}{\partial x} G(x,y) 
\Big\vert_{x = y +\epsilon}
-
\frac{\partial}{\partial x} G(x,y) 
\Big\vert_{x = y - \epsilon}
\right] = 1.
\end{equation}
Using (3.17), the partial derivative of $g(x,y;z,z')$ can be calculated as
\begin{align}
&\frac{\partial}{\partial x} g(x,y;z,z') 
\nonumber \\*
& \qquad =
\Biggl\{
\begin{array}{ll}
f g - ik N \langle \Phi_0, U(z,x) (Q_+ - Q_-) U(x,y) (Q_+ + Q_-) U(y,z') \Psi_0 \rangle, & \quad  x > y, 
\\[1ex]
f g - ik N \langle \Phi_0, U(z,y) (Q_+ + Q_-) U(y,x) (Q_+ - Q_-) U(x,z') \Psi_0 \rangle, & \quad y > x,
\end{array}
\end{align}
where $f g$ stands for $f(x) g(x,y;z,z')$.
Hence, 
\begin{multline}
\lim_{\epsilon \downarrow 0}
\left[
\frac{\partial}{\partial x} g(x,y;z,z') 
\Big\vert_{x = y +\epsilon}
- 
\frac{\partial}{\partial x} g(x,y;z,z') 
\Big\vert_{x = y - \epsilon}
\right] 
\\*
=2ikN(z,z') \langle \Phi_0, U(z,y) (Q_- Q_+ - Q_+ Q_-) U(y,z') \Psi_0 \rangle.
\end{multline}
It can be shown from (3.12), (3.13), and (2.2) that
\begin{equation}
(Q_- Q_+ - Q_+ Q_-) U = U (Q_- Q_+ - Q_+ Q_-).
\end{equation}
Also using  $Q_- \Psi_0 =Q_+^\dagger \Phi_0 = 0$ and $U(z,y) U(y,z')= U(z,z')$, we find
\begin{align}
\langle \Phi_0, U(z,y) (Q_- Q_+ - Q_+ Q_-) U(y,z') \Psi_0 \rangle
&=
\langle \Phi_0, U(z,z') Q_- Q_+  \Psi_0 \rangle
\nonumber \\*
&=
\langle \Phi_0,  Q_- Q_+ U(z,z') \Psi_0 \rangle.
\end{align}
As can be seen from (4.10) and (4.12), condition
(4.8) is satisfied if we choose $N(z,z')$ as
\begin{equation}
N(z,z')=\frac{1}{2 i k \langle \Phi_0, U(z,z') Q_- Q_+ \Psi_0 \rangle}.
\end{equation}
The denominator of (4.13) vanishes if $Q_+ \Psi_0 = 0$ or $Q_-^\dagger \Phi_0 = 0$ [see (4.12)], so it is necessary that $Q_+ \Psi_0 \neq 0$ and $Q_-^\dagger \Phi_0 \neq 0$.
From (4.6), (4.5), (4.3), and (4.13), we obtain the result
\begin{equation}
G(x,y) 
= 
\left\{
\begin{array}{ll}
\dfrac{\langle \Phi_0, U(\infty,x) (Q_+ + Q_-) U(x,y) (Q_+ + Q_-) U(y,-\infty) \Psi_0 \rangle}
 {2 i k \langle \Phi_0, U(\infty, -\infty) Q_- Q_+ \Psi_0 \rangle},
& \quad x \geq y,
\\[3ex]
\dfrac{\langle \Phi_0, U(\infty,y) (Q_+ + Q_-) U(y,x) (Q_+ + Q_-) U(x,-\infty) \Psi_0 \rangle}
{2 i k \langle \Phi_0, U(\infty, -\infty) Q_- Q_+ \Psi_0 \rangle},
& \quad y \geq x,
\end{array}
\right.
\end{equation}
where $\Psi_0$ and $\Phi_0$ are vectors such that
\begin{equation}
Q_- \Psi_0 = 0, \qquad Q_+^\dagger \Phi_0 = 0, \qquad
Q_+ \Psi_0 \neq 0, \qquad Q_-^\dagger \Phi_0 \neq 0.
\end{equation}
In the expressions on the right-hand side of (4.14), the denominator and the numerator do not exist separately.
The upper expression, for example, should be understood as shorthand for
\begin{equation}
\lim_{\mathstrut z \to +\infty} \lim_{\mathstrut z' \to -\infty} 
\frac{\langle \Phi_0, U(z,x) (Q_+ + Q_-) U(x,y) (Q_+ + Q_-) U(y,z') \Psi_0 \rangle}
 {2 i k \langle \Phi_0, U(z,z') Q_- Q_+ \Psi_0 \rangle}.
\end{equation}

For later use, let us slightly generalize (4.14).
Suppose that there are two, not only one, pairs of operators $Q_\pm$. 
That is to say, suppose that there are two pairs of operators, $Q_\pm^a$ and $Q_\pm^b$, each satisfying the commutation relations (3.7) in place of $Q_\pm$. 
We assume that the vectors $\Psi_0$ and $\Phi_0$ satisfy $Q_-^a \Psi_0=0$ and $(Q_+^b)^\dagger \Phi_0=0$. Then, we can replace (4.3) with
\begin{equation}
g_0(x,y;z,z')  \equiv \langle \Phi_0, U(z,x) (Q_+^b + Q_-^b) U(x,y) (Q_+^a + Q_-^a) U(y,z') \Psi_0 \rangle.
\end{equation}
Equations (4.4) -- (4.7) remain valid with this replacement.
Equations (4.10) -- (4.13) hold with $Q_- Q_+$ and $Q_+ Q_-$ replaced by $Q_-^b Q_+^a$ and $Q_+^b Q_-^a$, respectively. 
Thus, the Green function  can be written as
\begin{equation}
G(x,y) 
= 
\frac{\langle \Phi_0, U(\infty,x) (Q_+^b + Q_-^b) U(x,y) (Q_+^a + Q_-^a) U(y,-\infty) \Psi_0 \rangle}
 {2 i k \langle \Phi_0, U(\infty, -\infty) Q_-^b Q_+^a \Psi_0 \rangle}
 \qquad \mbox{for $x \geq y$}
\end{equation}
and similarly for $y \geq x$. The conditions on $\Psi_0$ and $\Phi_0$ are
\begin{equation}
Q_-^a \Psi_0 = 0, \qquad (Q_+^b)^\dagger \Phi_0=0, \qquad
Q_+^a \Psi_0 \neq 0, \qquad  (Q_-^b)^\dagger \Phi_0 \neq 0.
\end{equation}
Expression (4.14) can be thought of as a special case of (4.18), with $Q_\pm^b = Q_\pm^a$.

\section{$\boldsymbol{Q_+}$ and $\boldsymbol{Q_-}$ as elements of $\boldsymbol{sl(3,\mathbf{C})}$}
To incorporate $Q_\pm$ in addition to $J_\pm$ and $J_3$, we need to deal with a larger algebra which contains $sl(2,\mathbf{C})$ as a subalgebra. 
One such algebra is the Lie algebra $sl(3,\mathbf{C})$.
[Another possibility is the orthosymplectic Lie superalgebra $osp(1/2)$. See Appendix~E.] 
The Lie algebra $sl(3,\mathbf{C})$ is defined by the commutation relations
\begin{alignat}{6}
&[J_+, J_-] = 2 J_3, &\qquad 
&[K_+, K_-] = 2 K_3, &\qquad 
&[L_+, L_-] = 2 L_3, 
\nonumber \\
&[J_3, J_\pm] = \pm J_\pm, &\qquad 
&[K_3, K_\pm] = \pm K_\pm, &\qquad
&[L_3, L_\pm] = \pm L_\pm,
\nonumber \\
&[J_3, K_\pm] = \mp \tfrac{1}{2} K_\pm, &\qquad
&[K_3, L_\pm] = \mp \tfrac{1}{2} L_\pm, &\qquad
&[L_3, J_\pm] = \mp \tfrac{1}{2} J_\pm,
\nonumber \\
&[J_3, L_\pm] = \mp \tfrac{1}{2} L_\pm, &\qquad
&[K_3, J_\pm] = \mp \tfrac{1}{2} J_\pm, &\qquad
&[L_3, K_\pm] = \mp \tfrac{1}{2} K_\pm, 
\nonumber \\
&[J_\pm, K_\pm] = \pm L_\mp, &\qquad
&[K_\pm, L_\pm] = \pm J_\mp, &\qquad
&[L_\pm, J_\pm] = \pm K_\mp,
\nonumber \\
&[J_\pm, K_\mp] = 0, &\qquad
&[K_\pm, L_\mp] = 0, &\qquad
&[L_\pm, J_\mp] = 0,
\nonumber \\
&[J_3, K_3] = 0, &\qquad
&[K_3, L_3] = 0, &\qquad
&[L_3, J_3] = 0.
\end{alignat}
The consistency of (5.1) requires
\cite{Note4}
that
$J_3+K_3+L_3=0$.
The $sl(3,\mathbf{C})$ Lie algebra is spanned by eight independent elements $J_+$, $J_-$, $J_3$, $K_+$, $K_-$, $K_3$, $L_+$, and $L_-$ satisfying (5.1). The $sl(2,\mathbf{C})$  Lie algebra spanned by $J_+$, $J_-$, and $J_3$ is contained in $sl(3,\mathbf{C})$ as a subalgebra. 

We can see that the commutation relations of (3.7) are included in (5.1) if we identify
\begin{equation}
Q_+ = L_-, \quad Q_- = K_+ \qquad 
\text{or}
\qquad
Q_+ = -K_-, \quad Q_- = L_+.
\end{equation}
Unfortunately, we cannot use expression (4.14) with (5.2).
Since $[Q_+, Q_-]=0$ for either choice of (5.2), the denominator on the right-hand side of (4.13) becomes zero and so (4.14) does not make sense.  
This difficulty can be circumvented by adopting
\begin{equation}
Q_+ = c_1 L_- - c_2 K_-, \qquad Q_- = c_1 K_+ + c_2 L_+,
\end{equation}
where $c_1$ and $c_2$ are arbitrary nonzero numbers. 
However, since there are two possible choices for $Q_\pm$, it is more natural to use (4.18) rather than (4.14). We identify $Q_\pm^a$ and $Q_\pm^b$ as
\begin{equation}
Q_+^a= L_-, \qquad Q_-^a = K_+, \qquad
Q_+^b = - K_-, \qquad Q_-^b = L_+.
\end{equation}
Expression~(4.18) then becomes
\begin{equation}
G(x,y) 
= 
\frac{\langle \Phi_0, U(\infty,x) (L_+ - K_-) U(x,y) (L_- + K_+) U(y,-\infty) \Psi_0 \rangle}
 {2 i k \langle \Phi_0, U(\infty, -\infty) L_+ L_- \Psi_0 \rangle},
\end{equation}
where 
\begin{equation}
K_+ \Psi_0 =0, \qquad K_-^\dagger \Phi_0=0, 
\qquad
L_- \Psi_0 \neq 0, \qquad L_+^\dagger \Phi_0 \neq 0.
\end{equation}
[Here and hereafter, we consider $G(x,y)$ only for $x \geq y$. The expressions for $x < y$ can be obtained by interchanging $x$ and $y$.] 
Although we do not discuss it here, it can be shown that expression~(4.14) with (5.3) essentially reduces to (5.5) as well.

We are regarding $J_\pm$, $J_3$, $K_\pm$, $K_3$, $L_\pm$, and $L_3$ as linear operators (or matrices) acting on some vector space. 
This means that we are actually dealing with a representation of the Lie algebra $sl(3,\mathbf{C})$. 
Choosing an explicit form for these operators is equivalent to choosing a particular representation of $sl(3,\mathbf{C})$. 
When a particular representation of the Lie algebra is chosen, the evolution operator $U$ defined by (3.1) belongs to the corresponding representation of the Lie group $SL(3,\mathbf{C})$.
Equation~(5.5) is a representation-independent expression of the Green function.
By writing (5.5) in specific representations, we obtain diverse expressions for the Green function. We will see examples in Secs.~VI and VII.

\section{Fundamental representation}

The fundamental representation of the Lie algebra $sl(3,\mathbf{C})$ consists of $3\times 3$ traceless matrices. We can choose the $3 \times 3$ traceless matrices satisfying (5.1) as
\begin{alignat}{5}
J_3 &=
\frac{1}{2}
\begin{pmatrix}
-1 &  0 \,& 0 \,\\
0 & 1 \,& 0 \,\\
0 & 0 \,& 0 \,
\end{pmatrix},
&\qquad
J_+ &=
\begin{pmatrix}
0 & 0 \,& 0 \,\\
-1 & 0 \,& 0 \,\\
0 & 0 \,& 0 \,
\end{pmatrix},
&\qquad
J_- &=
\begin{pmatrix}
\,0 & -1 & 0 \,\\
\,0 & 0 & 0 \,\\
\,0 & 0 & 0 \,
\end{pmatrix},
\nonumber \\[1ex]
K_3 &=
\frac{1}{2}
\begin{pmatrix}
\,0 &  0 & 0 \,\\
\,0 & -1 & 0 \,\\
\,0 & 0 & 1 \,
\end{pmatrix}, 
&\qquad
K_+ &=
\begin{pmatrix}
\,0 & 0 & 0 \,\\
\,0 & 0 & 0 \,\\
\,0 & -1 & 0 \,
\end{pmatrix},
&\qquad
K_- &=
\begin{pmatrix}
\,0 & \,0 & 0 \\
\,0 & \,0 & -1 \\
\,0 & \,0 & 0
\end{pmatrix},
\nonumber \\[1ex]
L_3 &=
\frac{1}{2}
\begin{pmatrix}
\,1 & \,0 & 0 \\
\,0 & \,0 & 0 \\
\,0 & \,0 & -1
\end{pmatrix}, 
&\qquad
L_+ &=
\begin{pmatrix}
\,0 & \,0 & -1 \\
\,0 & \,0 & 0 \\
\,0 & \,0 & 0
\end{pmatrix},
&\qquad
L_-&=
\begin{pmatrix}
0 & 0 \,& 0 \,\\
0 & 0 \,& 0 \,\\
-1 & 0 \,& 0 \,
\end{pmatrix}.
\end{alignat}
The representation space, i.e., the vector space on which these matrices act, is the space of three-dimensional vectors with complex components. The inner product in this space is
\begin{equation}
\langle \Phi, \Psi \rangle = a_1^*\,b_1^{\mathstrut} + a_2^*\, b_2^{\mathstrut} + a_3^*\,b_3^{\mathstrut}
\qquad
\mbox{for}
\qquad
\Phi = 
\begin{pmatrix}
\,a_1\, \\
a_2 \\
a_3
\end{pmatrix},
\quad
\Psi = 
\begin{pmatrix}
\,b_1\, \\
b_2 \\
b_3
\end{pmatrix}.
\end{equation}
With this inner product, the adjoint is the conjugate transpose of the matrix, so
$K_-^\dagger = K_+$ and $L_+^\dagger = L_-$.
The vectors satisfying (5.6) can be chosen as
\begin{equation}
\Psi_0 = \Phi_0 =
\begin{pmatrix}
\,1 \,\\
0 \\
0 \\
\end{pmatrix}.
\end{equation}
Since the upper-left $2 \times 2$ parts of $J_3$ and $J_1 = (J_+ + J_-)/2 $ coincide with the $2 \times 2$ matrices of (1.9), 
it is obvious that $U$ defined by (3.1) is a matrix of the form\begin{equation}
U(x_2,x_1) = 
\begin{pmatrix}
\alpha(x_2,x_1;k) & \beta(x_2,x_1;-k) & \, \, 0 \, \, \\
\beta(x_2,x_1;k) & \alpha(x_2,x_1;-k) & \, \, 0  \, \,\\
0 & 0 & \, \, 1 \, \,
\end{pmatrix},
\end{equation}
where $\alpha$ and $\beta$ are the same functions as in (1.7).
This $U$ is a $3 \times 3$ matrix with unit determinant and so belongs to the fundamental representation of the $SL(3,\mathbf{C})$ Lie group.
Substituting (6.4), (6.1), and (6.3) in (5.5) yields
\begin{equation}
G(x,y)=
\frac{[\alpha(\infty,x;k) - \beta(\infty,x;-k)][\alpha(y,-\infty;k) + \beta(y, -\infty;k)]}
{2ik \alpha(\infty, -\infty;k)}.
\end{equation}
As noted before, the denominator and the numerator on the right-hand side of (6.5) do not exist separately. 
The numerator is $\psi_+(x,\infty) \psi_-(y,-\infty)$ [see (2.4) and (1.8)], which is meaningless by itself. 
We need to normalize $\psi_\pm$ as in Eqs.~(3.23), which can be written as
\begin{equation}
\phi_+(x) =
\frac{\alpha(\infty,x;k) - \beta(\infty,x;-k)}{\alpha(\infty,x_0;k)},
\qquad
\phi_-(x) =
\frac{\alpha(x,-\infty;k) + \beta(x,-\infty;k)}{\alpha(x_0,-\infty;k)}.
\end{equation}
From (6.6), (1.6), and (2.3), we have 
\begin{equation}
\frac{d\phi_+}{dx\,} = f \phi_+ 
+ i k \frac{\alpha(\infty,x;k) + \beta(\infty,x;-k)}{\alpha(\infty,x_0;k)}, 
\qquad
\frac{d\phi_-}{dx\,} = f \phi_- 
- i k \frac{\alpha(x,-\infty;k) - \beta(x,-\infty;k)}{\alpha(x_0,-\infty;k)}.
\end{equation}
Hence we can calculate the Wronskian $W(\phi_+, \phi_-) 
\equiv \phi_+ (d\phi_- /dx)- (d\phi_+/dx)\phi_- $as
\begin{align}
W(\phi_+, \phi_-) 
&=
- 2 i k 
\frac{\alpha(\infty,x;k) \alpha(x,-\infty;k)
+\beta(\infty,x;-k) \beta(x,-\infty;k)}
{\alpha(\infty,x_0;k) \alpha(x_0,-\infty;k)}
\nonumber \\*
&=
\frac{-2 i k \alpha(\infty, -\infty;k)}
{\alpha(\infty,x_0;k) \alpha(x_0,-\infty;k)},
\end{align}
where we have used (2.1).
From (6.5), (6.6) and (6.8), we retrieve the well-known expression\begin{equation}
G(x,y)=
- \frac{\phi_+(x) \phi_-(y)}
{W(\phi_+, \phi_-)}.
\end{equation}

\section{Infinite-dimensional representation}
It can be readily seen that the following differential operators, which act on functions of two variables $\xi$ and $\mu$, satisfy the $sl(2,\mathbf{C})$ commutation relations (3.5):
\begin{equation}
J_+ = -\xi^2 \frac{\partial}{\partial \xi} - \xi \mu \frac{\partial}{\partial \mu}, 
\qquad
J_- = \frac{\partial}{\partial \xi}, 
\qquad
J_3= \xi \frac{\partial}{\partial \xi} + \frac{1}{2} \mu \frac{\partial}{\partial \mu}.
\end{equation}
This representation of $sl(2,\mathbf{C})$ was thoroughly studied in Ref.~\onlinecite{formulation}, where it was used for the analysis of reflection coefficients.
[The meaning of (7.1) will be explained in Sec.~X.]

We can extend (7.1) to $sl(3,\mathbf{C})$ in the following way.
Interchanging $\xi$ and $\eta$ in (7.1) and changing the overall sign, we define
$L_-$, $L_+$, and $L_3$ as
\begin{equation}
L_- \equiv \mu^2 \frac{\partial}{\partial \mu} + \mu \xi \frac{\partial}{\partial \xi}, 
\qquad
L_+ \equiv - \frac{\partial}{\partial \mu}, 
\qquad
L_3 \equiv - \mu \frac{\partial}{\partial \mu} - \frac{1}{2} \xi \frac{\partial}{\partial \xi}.
\end{equation}
It is obvious that $L_\pm$ and $L_3$ defined by (7.2) also satisfy the $sl(2,\mathbf{C})$ commutation relations. 
From (7.1) and (7.2), we define 
$K_+ \equiv [J_-, L_-] = \mu \partial/\partial \xi$, 
$K_- \equiv [L_+, J_+] = \xi \partial/\partial \mu$, 
and $K_3 \equiv  -J_3 - L_3 = \frac{1}{2} \mu (\partial/\partial \mu)  - \frac{1}{2} \xi (\partial/\partial \xi)$. 
In sum, we have
\begin{gather}
\!\!\!\!\!\!\!
J_+ = -\xi^2 \frac{\partial}{\partial \xi} - \xi \mu \frac{\partial}{\partial \mu}, 
\qquad \ 
J_- = \frac{\partial}{\partial \xi}, 
\qquad \ 
J_3= \xi \frac{\partial}{\partial \xi} + \frac{1}{2} \mu \frac{\partial}{\partial \mu},
\nonumber
\\
K_+ = \mu \frac{\partial}{\partial \xi}, 
\qquad \ \ \ \ \ \ 
K_- = \xi \frac{\partial}{\partial \mu}, 
\qquad \ \ \ \ \ \  
K_3 = \frac{1}{2} \mu \frac{\partial}{\partial \mu} - \frac{1}{2} \xi \frac{\partial}{\partial \xi},
\nonumber
\\
L_+ = -\frac{\partial}{\partial \mu}, 
\qquad
L_- = \mu^2 \frac{\partial}{\partial \mu} + \mu \xi \frac{\partial}{\partial \xi}, 
\qquad
L_3= -\mu \frac{\partial}{\partial \mu} - \frac{1}{2} \xi\frac{\partial}{\partial \xi}.
\end{gather}
It can be checked that all commutation relations of (5.1) are satisfied with (7.3). 
Thus, (7.3) gives a representation of $sl(3,\mathbf{C})$. 
The meaning of these operators, as well as the meaning of the variables $\mu$ and $\xi$, will be discussed in Sec.~X.

The representation space, on which these operators act, is a vector space consisting of functions of two complex variables $\xi$ and $\mu$. We assume that these functions are analytic in $\vert \xi \vert < 1$ and $\vert \mu \vert < 1$. 
The basis of the representation space is given by
\begin{equation}
\Psi_{p,q}(\xi,\mu) \equiv \xi^p \mu^q,
\qquad
p=0,1,2,\ldots, \quad q=0,1,2,\ldots.
\end{equation}
We define the inner product of the basis vectors as
\begin{equation}
\langle \Psi_{p'\!,q'}, \Psi_{p,q} \rangle 
\equiv \delta_{p p'} \delta_{q q'}  \frac{p! \, q!}{(p+q-1)!},
\end{equation}
where $\delta_{m n}$ is the Kronecker delta.
[For $p=q=0$, we interpret (7.5) to mean $\langle \Psi_{0,0}, \Psi_{0,0} \rangle =0$. This inner product is not positive definite.]
Then it can be shown (see Appendix~F) that
\begin{gather}
J_\pm^\dagger = - J_\mp^{\mathstrut}, 
\qquad \!  
K_\pm^\dagger = K_\mp^{\mathstrut}, 
\qquad \!
L_\pm^\dagger = - L_\mp^{\mathstrut},
\qquad \!
J_3^\dagger =J_3^{\mathstrut}, 
\qquad \! 
K_3^\dagger = K_3^{\mathstrut}, 
\qquad \!
L_3^\dagger = L_3^{\mathstrut}.
\end{gather}
The vectors satisfying (5.6) can be chosen as
\begin{equation}
\Psi_0 = \Phi_0= \Psi_{0,q} \quad (q \neq 0),
\end{equation}
where $q$ is arbitrary as long as $q\neq 0$. 
The inner product of arbitrary two functions $\Psi(\xi, \mu)$ and $\Phi(\xi, \mu)$ is determined from (7.5) by the linearity of the inner product in the right entry and the conjugate linearity in the left entry.
We have (see Appendix~F) 
\begin{multline}
\langle \Phi, \Psi \rangle 
=
\frac{1}{\pi^2}
\iint_{\vert \xi \vert < 1} \! \! \! d\xi_1 d \xi_2\,
\iint_{\vert \mu \vert < 1} \! \! \! d\mu_1 d \mu_2 \,
\delta(1 - \vert \xi \vert ^2 - \vert \mu \vert^2)
\\*
\times
\left[ \Phi(\xi,\mu)\right]^*
\left(1 + \xi \frac{\partial}{\partial \xi} + \mu \frac{\partial}{\partial \mu} \right)
\left( \xi \frac{\partial}{\partial \xi} + \mu \frac{\partial}{\partial \mu} \right)
\Psi(\xi,\mu),
\end{multline}
where $\xi = \xi_1 + i \xi_2$, $\mu = \mu_1 + i \mu_2$.
The integrals in (7.8) are area integrals over the unit disk $\vert \xi \vert < 1$ in the complex $\xi$-plane and $\vert \mu \vert < 1$ in the $\mu$-plane.

It was shown in Ref.~\onlinecite{formulation} that the evolution operator $U$, which is given by (3.6) with (7.1), acts on an arbitrary function $\Psi(\xi,\mu)$ as
\begin{equation}
U(x_2,x_1) \Psi(\xi, \mu) = \Psi(\hat{L}(x_2,x_1;\xi), \hat{T}(x_2,x_1;\xi,\mu)),
\end{equation}
where $\hat L$ and $\hat T$ are defined by
\begin{equation}
\hat{L}(x_2,x_1;\xi) \equiv R_l(x_2,x_1) + \frac{\xi\, [\tau(x_2,x_1)]^2}{1 - \xi R_r(x_2,x_1)},
\qquad
\hat{T}(x_2,x_1;\xi,\mu) \equiv \frac{\mu \tau(x_2,x_1)}{1 - \xi R_r(x_2,x_1)}.
\end{equation}
Although Eq.~(7.9) was derived in Ref.~\onlinecite{formulation} for $SL(2,\mathbf{C})$, it can also be used for $SL(3,\mathbf{C})$ without need of any modification. 
We can think of this $U$ as an operator belonging to the infinite-dimensional representation of the Lie group $SL(3,\mathbf{C})$ corresponding to (7.3).

Let us consider (5.5) in this representation. We use the abbreviation
\begin{align}
U(3) \equiv U(z,x), \qquad  \ \ \ \ 
U(2) &\equiv U(x,y), \qquad  \ \ \ \ 
U(1) \equiv U(y,z'), 
\nonumber \\
\hat{T}(3;\xi,\mu) \equiv \hat{T}(z,x;\xi,\mu),
\qquad \! \! \!
\hat{T}(2;\xi,\mu) &\equiv \hat{T}(x,y;\xi,\mu), 
\qquad \! \! \!
\hat{T}(1;\xi,\mu) \equiv \hat{T}(y,z';\xi,\mu), 
\end{align}
and similarly for $\tau$, $R_r$, $R_l$, and $\hat{L}$. 
With (7.7), expression (5.5)  reads
\begin{equation}
2 i k G(x,y) 
= 
\lim_{\mathstrut z \to \infty} \lim_{\mathstrut z' \to -\infty}
\frac{\langle \Psi_{0,q}, U(3) (L_+ - K_-) U(2) (L_- + K_+) U(1) \Psi_{0,q} \rangle}
 {\langle \Psi_{0,q}, U(z, z') L_+ L_- \Psi_{0,q} \rangle},
\end{equation}
where $q$ is an arbitrary positive integer (the result will not depend on $q$). 
Now we calculate the numerator on the right-hand side of  of (7.12) using (7.9). 
First, (7.9) and (7.10) give
\begin{equation}
U(1) \Psi_{0,q} 
= U(1) \mu^q
= \bigl[\hat{T}(1;\xi,\mu)\bigr]^q
=\left[
\frac{\mu \tau(1)}{1 - \xi R_r(1)} 
\right]^q.
\end{equation}
Next, we apply $L_- + K_+$ to (7.13). 
Here we use the formula
\begin{equation}
(L_- + K_+) 
\left(
\frac{\mu}{1 - c \xi}
\right)^m
= m (1 + c)
\left(
\frac{\mu}{1 - c \xi}
\right)^{m+1},
\end{equation}
which holds for any positive integer $m$ and any complex number $c$.
This formula can be checked by direct calculation with 
$L_- + K_+ = \mu^2 (\partial/\partial \mu) + \mu (1 + \xi) (\partial/\partial \xi)$.
Hence,
\begin{equation}
(L_- + K_+) U(1) \Psi_{0,q} 
= q
\left[
\frac{\mu}{1 - \xi R_r(1)} 
\right]^{q + 1}
 [\tau(1)]^q [1 + R_r(1)].
\end{equation}
Using (7.9), (7.10), and $L_+ - K_- = - (1 + \xi) (\partial/\partial \mu)$, we can proceed as follows:
\begin{multline}
U(2) (L_- + K_+)   U(1) \Psi_{0,q} 
= q 
\left[
\frac{\hat{T}(2;\xi,\mu)}{1 - \hat{L}(2;\xi) R_r(1)} 
\right]^{q + 1}
[\tau(1)]^q [1 + R_r(1)]
\\*
=
q 
\frac{
\mu^{q+1} 
[\tau(2)]^{q+1} [\tau(1)]^q [1 + R_r(1)]
}{
\bigl\{
[1 - \xi R_r(2)][1 - R_l(2) R_r(1)] - \xi [\tau(2)]^2 R_r(1)]
\bigr\}^{q+1}
},
\end{multline}
\vspace{-25pt}
\begin{multline}
(L_+ - K_-)U(2) (L_- + K_+) U(1) \Psi_{0,q} 
\\*
=
- q (q+1) 
\frac{
(1 + \xi)\,\mu^q 
[\tau(2)]^{q+1} [\tau(1)]^q [1 + R_r(1)]
}{
\bigl\{
[1 - \xi R_r(2)][1 - R_l(2) R_r(1)] - \xi [\tau(2)]^2 R_r(1)]
\bigr\}^{q+1}
},
\end{multline}
\vspace{-25pt}
\begin{multline}
U(3)(L_+ - K_-)U(2) (L_- + K_+) U(1) \Psi_{0,q} 
\\*
=
-q (q+1)  
\frac{
\bigl[1 + \hat{L}(3;\xi)\bigr] \bigl[\hat{T}(3;\xi,\mu)\bigr]^q
[\tau(2)]^{q+1} 
[\tau(1)]^q
[1 + R_r(1)]}{
\bigl\{
[1 - \hat{L}(3;\xi) R_r(2)]
[1 - R_l(2) R_r(1)]
 - \hat{L}(3;\xi) [\tau(2)]^2 R_r(1)
\bigr\}^{q+1}
}.
\end{multline}
The right-hand side of (7.18) has the form of $\Phi(\xi,\mu)=[\hat T(\xi,\mu)]^q A(\hat L(\xi))=\mu^q B(\xi)$, where $A$ is a function of $\hat L$ and $B$ is a function of $\xi$. 
We find from (7.5) that
\begin{equation}
\langle \Psi_{0,q}, \Phi \rangle = q B(0)
\qquad
\text{for} 
\quad \Phi(\xi,\mu) = \mu^q B(\xi).
\end{equation}
Since $\hat T(\xi = 0,\mu) = \mu \tau$ and $\hat L(\xi=0) = R_l$, we have $q B(0) = q \tau^q A(R_l)$.
Therefore, the inner product of (7.18) with $\Psi_{0,q}$ is obtained by replacing $\hat T$ and $\hat L$ with $\tau$ and $R_l$, respectively, and then multiplying by~$q$.
This yields
\begin{align}
\langle \Psi_{0,q}, U(3) & (L_+ - K_-)U(2) (L_- + K_+) U(1) \Psi_{0,q} \rangle
\nonumber \\*
&=
-q^2 (q+1) 
\frac{
[1 + R_l(3)] \, \tau(2) \, [1 + R_r(1)] [\tau(3)\tau(2) \tau(1)]^q
}{
\bigl\{
[1 - R_l(3) R_r(2)][1 - R_l(2) R_r(1)] - R_l(3) [\tau(2)]^2 R_r(1)]
\bigr\}^{q+1}
}
\nonumber \\*
&=
-q^2 (q+1) 
\frac{
[1 + R_l(3)] \, \tau(2) \, [1 + R_r(1)] [\tau(z,z')]^q
}{
[1 - R_l(3) R_r(2)][1 - R_l(2) R_r(1)] - R_l(3) [\tau(2)]^2 R_r(1)]
},
\end{align}
where we have used
\begin{equation}
\tau(z,z')
=
\frac{
\tau(3)\tau(2)\tau(1) 
}{
[1 - R_l(3) R_r(2)][1 - R_l(2) R_r(1)] - R_l(3) [\tau(2)]^2 R_r(1)
},
\end{equation}
the proof of which is given in Appendix~G.

In the meantime, the denominator on the right-hand side of (7.12) can be easily calculated using (7.9) and (7.19) as
\begin{gather}
L_+ L_- \Psi_{0,q} = - q (q+1) \mu^q, \qquad
U(z,z') L_+ L_- \Psi_{0,q} = -q (q+1) [\hat{T}(z,z';\xi,\mu)]^q,
\\
\langle \Psi_{0,q}, U(z,z') L_+ L_- \Psi_{0,q} \rangle 
= -q^2 (q+1) [\tau(z,z')]^q.
\end{gather}
Substituting (7.20) and (7.23) into (7.12), we obtain 
\begin{multline}
2 i k G(x,y)
\\*
=
\frac{
[1 + R_l(\infty,x)] \, \tau(x,y) \,[1 + R_r(y,-\infty)]
}{
[1 - R_l(\infty,x) R_r(x,y)][1 - R_l(x,y) R_r(y,-\infty)] - R_l(\infty,x) [\tau(x,y)]^2 R_r(y,-\infty)
}.
\end{multline}
Unlike (6.5), both the denominator and the numerator of (7.24) exist as finite quantities.

Thus, in this representation, Eq.~(5.5) gives an expression of the Green function in terms of transmission and reflection coefficients. 
Solutions of the Schr\"odinger equation can be interpreted as waves undergoing the process of multiple reflections and transmissions.
The infinite-dimensional representation given by (7.3) provides a natural basis for understanding  the Schr\"odinger equation from this viewpoint, as we will see in Sec.~X.

\section{More general formulas}
As shown in (5.2), the Lie algebra $sl(3,\mathbf{C})$ includes two different pairs of $Q_\pm$ satisfying (3.7). All equations derived in Sec.~III hold for both choices of (5.2). Equations (3.11) read
\begin{subequations}
\begin{alignat}{3}
U L_- &= \tau L_- U + R_l U K_+, &\qquad K_+ U &= \tau U K_+ + R_r L_- U, \\
U K_- &= \tau K_- U - R_l U L_+, &\qquad L_+ U &= \tau U L_+ - R_r K_- U.
\end{alignat}
\end{subequations}
Let $\Psi_0$ and $\Phi_0$ be vectors satisfying $K_+ \Psi_0 =0$ and $K_-^\dagger \Phi_0=0$. The two equations of (8.1a) give 
$ U L_- \Psi_0 = \tau L_- U \Psi_0 $ and $K_+ U \Psi_0 = R_r L_- U \Psi_0$, 
and hence $(L_- + K_+) U \Psi_0 
= (1/\tau)(1 + R_r) U L_- \Psi_0$.
Repeating this $n$~times, and also using $[K_+, L_-]=0$, we have
\begin{equation}
(L_- + K_+)^n U \Psi_0 
=
\left(
\frac{1 + R_r}{\tau}
\right)^n
U L_-^n \Psi_0.
\end{equation}
In the same way, by using (8.1b) and $[K_-, L_+]=0$, we can derive
\begin{equation}
\langle \Phi_0, U (L_+ - K_-)^n \,\cdots  \, \rangle
=
\left(
\frac{1 + R_l}{\tau}
\right)^n
\langle \Phi_0, L_+^n U \,\cdots \, \rangle.
\end{equation}
The second equation of (8.1b) also gives
\begin{equation}
\langle \Phi_0, L_+^n U \, \cdots \, \rangle
= \tau^n \langle \Phi_0, U L_+^n \, \cdots \, \rangle.
\end{equation}
From (8.2) -- (8.4), we have
\begin{align}
\langle \Phi_0, U(z,x) (L_+ & - K_-)^n U(x,y) (L_- + K_-) ^n U(y,z') \Psi_0 \rangle
\nonumber \\*
&=
\left[
\frac{1 + R_l(z,x)}{\tau(z,x)}
\right]^n
\left[
\frac{1 + R_r(y,z')}{\tau(y,z')}
\right]^n
\langle \Phi_0, L_+^n U(z,z') L_-^n \Psi_0 \rangle
\nonumber \\*
&=
\left[
\frac{1 + R_l(z,x)}{\tau(z,x)}
\frac{1 + R_r(y,z')}{\tau(y,z')}
\,\tau(z,z')
\right]^n
\langle \Phi_0, U(z,z') L_+^n L_-^n \Psi_0 \rangle.
\end{align}
For this expression to be nonzero, it is necessary that $L_-^n \Psi_0 \neq 0$ and $(L_+^\dagger)^n \Phi_0 \neq 0$.
Setting $n=1$ in (8.5), and comparing it with (5.5), we find
\begin{equation}
2ik G(x,y) 
=
\lim_{\mathstrut z \to +\infty} \lim_{\mathstrut z' \to -\infty}
\frac{1 + R_l(z,x)}{\tau(z,x)}
\frac{1 + R_r(y,z')}{\tau(y,z')}
\,\tau(z,z').
\end{equation}
[This is the same equations as (6.5).] From (8.5) and (8.6), we obtain
\begin{equation}
[2 i k G(x,y)]^n
=
\frac{
\langle \Phi_0, U(\infty,x) (L_+ - K_-)^n U(x,y) (L_- + K_-) ^n U(y,-\infty) \Psi_0 \rangle
}{
\langle \Phi_0, U(\infty,-\infty) L_+^n  L_-^n \Psi_0 \rangle
}
\end{equation}
for any positive integer~$n$. The vectors $\Psi_0$ and $\Phi_0$ in (8.7) are required to satisfy
\begin{equation}
K_+ \Psi_0 =0, \qquad K_-^\dagger \Phi_0=0, 
\qquad
L_-^n \Psi_0 \neq 0, \qquad (L_+^\dagger)^n \Phi_0 \neq 0.
\end{equation}

In the fundamental representation studied in Sec.~VI, vectors satisfying (8.8) with (6.1) do not exist for $n \geq 2$.
In the representation introduced in Sec.~VII, on the other hand, the vector $\Psi_{0,q}$ [defined by (7.4)] satisfies (8.8) with (7.3) for any $n$ if $q\neq 0$. 
Let us check Eq.~(8.7) in this representation. 
The $n$-times repetition of (7.14) gives
\begin{equation}
(L_- + K_+)^n 
\left(
\frac{\mu}{1 - c \xi}
\right)^m
=\frac{(m + n - 1)!}{(m - 1)!}\,
(1 + c)^n
\left(
\frac{\mu}{1 - c \xi}
\right)^{m+n}.
\end{equation}
From (7.13) and (8.9) we have, with the notation of (7.11), 
\begin{align}
(L_- + K_+)^n U(1) \Psi_{0,q}
= \frac{(q+n-1)!}{(q-1)!}
\left[
\frac{\mu}{1 - \xi R_r(1)}
\right]^{q+n}
[\tau(1)]^q
[1 + R_r(1)]^n.
\end{align}
Proceeding in the same way as in Eqs.~(7.16) -- (7.20), we arrive at
\begin{align}
&\langle \Psi_{0,q}, U(3) (L_+ - K_-)^n U(2) (L_- + K_+)^n U(1) \Psi_{0,q} \rangle
\nonumber \\*
& \qquad 
= (-1)^n q \,\frac{(q+n)!}{q!}\frac{(q+n-1)!}{(q-1)!}\,
\nonumber \\*
& \qquad \qquad 
\times
\left\{
\frac{
[1 + R_l(3)] \tau(2) [1 + R_r(1)]
}{
[1 - R_l(3) R_r(2)] [1 - R_l(2) R_r(1)] - R_l(3) [\tau(2)]^2 R_r(1)
}
\right\}^n
\bigl[\tau(z,z')\bigr]^q.
\end{align}
As a generalization of (7.23), we can easily derive
\begin{equation}
\langle \Psi_{0,q}, U(z,z') L_+^n L_-^n \Psi_{0,q} \rangle
= 
(-1)^n q \, \frac{(q+n)!}{q!}\frac{(q+n-1)!}{(q-1)!}\,
[\tau(z,z')]^q.
\end{equation}
Hence,
\begin{multline}
\frac{
\langle \Psi_{0,q}, U(3) (L_+ - K_-)^n U(2) (L_- + K_+)^n U(1) \Psi_{0,q} \rangle
}{
\langle \Psi_{0,q}, U(z,z') L_+^n L_-^n \Psi_{0,q} \rangle
}
\\*
=
\left\{
\frac{
[1 + R_l(3)] \, \tau(2) \, [1 + R_r(1)]
}{
[1 - R_l(3) R_r(2)][1 - R_l(2) R_r(1)] - R_l(3) [\tau(2)]^2 R_r(1)
}
\right\}^n.
\end{multline}
This becomes (8.7) by letting $z \to \infty$ and $z' \to -\infty$ [see (7.24)].

Expression~(8.7) can be generalized to the product of Green functions with different arguments. 
For instance, the product of two Green functions and the product of three Green functions can be expressed as
\begin{subequations}
\begin{align}
& (2ik)^2 G(x_2,y_2) G(x_1,y_1) 
=
N_2
\langle \Phi_0, 
U(\infty,x_2) (L_+ - K_-) U(x_2,x_1) (L_+ - K_-) U(x_1,y_2) 
\nonumber \\*
& \qquad \qquad \qquad \qquad \qquad \qquad \quad \times
(L_- + K_+) U(y_2,y_1) (L_- + K_+) U(y_1,-\infty) \Psi_0 \rangle,
\\
& (2ik)^3 G(x_3,y_3) G(x_2,y_2) G(x_1,y_1)
\nonumber \\*
&\quad 
=
N_3
\langle \Phi_0, 
U(\infty,x_2) (L_+ - K_-) U(x_3,x_2) (L_+ - K_-) U(x_2,x_1) (L_+ - K_-) U(x_1,y_3) 
\nonumber \\* 
& \qquad \qquad \quad\times
 (L_- + K_+) U(y_3,y_2) (L_- + K_+) U(y_2,y_1) (L_- + K_+) U(y_1,-\infty) \Psi_0 \rangle,
\end{align}
\end{subequations}
\begin{equation}
N_2 \equiv  
\frac{1}
 {\langle \Phi_0, U(\infty, -\infty) L_+^2 L_-^2 \Psi_0 \rangle},
\qquad 
N_3 \equiv 
\frac{1}
 {\langle \Phi_0, U(\infty, -\infty) L_+^3 L_-^3 \Psi_0 \rangle}.
\end{equation}
[Equations (8.14) are to be understood together with (8.15), as in (4.16). The quantities $N_2$ and $N_3$ do not have a meaning by themselves.]
The vectors $\Psi_0$ and $\Phi_0$ are required to satisfy (8.8) with $n=2$ for (8.14a) and $n=3$ for (8.14b). 
We can easily prove (8.14) in the same way as (8.7), by using repeatedly (8.2) and (8.3) with $n=1$ and also using $[K_+, L_-]=[K_-,L_+]=0$.
In (8.14), we are assuming (without loss of generality) that $x_i \geq y_i$ for each $i$. 
The order between $x_i$ and $y_j$ with $i \neq j$ is arbitrary. 
[For example, (8.14a) is valid for either $x_1 \geq y_2$ or $y_2 > x_1$. 
Recall that $U(x_1,y_2)$ with $y_2>x_1$ is the inverse of $U(y_2,x_1)$.] 
The product of four or more Green functions can be expressed in the same way.
Products of Green functions frequently appear in perturbation expansions. Expressions like (8.14) can be useful for the perturbative treatment of the Schr\"odinger equation.

We cam also extend (8.7) to negative powers of $G$.
Let us assume that the operators $L_- + K_+$, $L_+ - K_-$, and $L_\pm$ have inverses denoted by $(L_- + K_+)^{-1}$, $(L_+ - K_-)^{-1}$, and $L_\pm^{-1}$.
(Although the matrices for $L_- + K_+$, $L_+ - K_-$, and $L_\pm$ in finite-dimensional representations are not invertible, it is possible to define the inverses of these operators in infinite-dimensional representations, as shown below.)
We define $A^{-n} \equiv (A^{-1})^n$ for any operator $A$.
Then,
\begin{equation}
\frac{1}{[2 i k G(x,y)]^n}
=
\frac{
\langle \Phi_0, U(\infty,x) (L_+ - K_-)^{-n} U(x,y) (L_- + K_+) ^{-n} U(y,-\infty) \Psi_0 \rangle
}{
\langle \Phi_0, U(\infty,-\infty) L_+^{-n} L_-^{-n} \Psi_0 \rangle
},
\end{equation}
where $\Psi_0$ and $\Phi_0$ are nonzero vectors such that $L_-^{-n} \Psi_0$ and $(L_+^{-n})^\dagger \Phi_0$ exist. As always, we also assume $K_+ \Psi_0 =0$ and $K_-^\dagger \Phi_0=0$. 
The proof of (8.16) is given in Appendix~H.

Let us check the validity of (8.16) in the representation given by (7.3).
We are assuming that the operators defined by (7.3) act on functions $\Psi(\xi,\mu)$ which are analytic in $\vert \xi \vert < 1$ and $\vert \mu \vert < 1$. 
We further restrict the domains of $L_- + K_+$, $L_+ - K_-$, and $L_\pm$ to functions $\Psi(\xi,\mu)$ satisfying $\Psi(\xi,0)=0$.
Then, obviously, the inverses of $L_+ = - \partial/\partial \mu$ and $L_+ - K_- = -(1 + \xi) \partial/\partial \mu$ are
\begin{equation}
L_+^{-1} \Phi(\xi,\mu) = -\int_0^\mu  \Phi(\xi,\mu) \,d\mu, \qquad
(L_+ - K_-)^{-1} \Phi(\xi,\mu) = - \frac{1}{1 + \xi}\int_0^\mu  \Phi(\xi,\mu) \,d\mu.
\end{equation}
The inverses of $L_-$ and $L_- + K_+$ are more complicated. 
We consider their action on functions of the form $\mu^m  g(\xi)$ instead of general $\Phi(\xi,\mu)$. It can be shown (see Appendix~I) that 
\begin{subequations}
\begin{align}
L_-^{-1} \mu^m g(\xi) 
&= \mu^{m-1} \frac{1}{\xi^{m-1}}
\int_0^\xi \xi^{m-2} g(\xi)\,d\xi,
\\
(L_- + K_+)^{-1} \mu^m g(\xi) 
&= \mu^{m-1} \frac{1}{(1 + \xi)^{m-1}}
\int_{-1}^\xi (1 + \xi)^{m-2} g(\xi)\,d\xi,
\end{align}
\end{subequations}
where $m > 1$. 
We can choose $\Psi_0$ and $\Phi_0$ in (8.16) as 
$\Psi_0 = \Phi_0 = \Psi_{0,q}$ [Eq.~(7.4)] with arbitrary $q>n$.
To calculate the numerator on the right-hand side of (8.16), we start with
\begin{equation}
(L_- + K_+)^{-n} U(1) \Psi_{0,q} = (L_- + K_+)^{-n} 
\left[
\frac{\mu \tau(1)}{1 - \xi R_r(1)}
\right]^q.
\end{equation}
As shown in Appendix~I, the inverse operator $(L_- + K_+)^{-1}$ given by (8.18b) satisfies
\begin{equation}
(L_- + K_+)^{-n}
\left(
\frac{\mu}{1 - c \xi}
\right)^m
=
\frac{(m-n-1)!}{(m-1)!}
\frac{1}{(1 + c)^n}
\left(
\frac{\mu}{1 - c \xi}
\right)^{m-n}
\end{equation}
for any positive integer $n$ smaller than $m$.
Therefore, (8.19) becomes
\begin{align}
(L_- + K_+)^{-n} U(1) \Psi_{0,q}
=
\frac{(q-n-1)!}{(q-1)!}
\left[
\frac{\mu}{1 - \xi R_r(1)}h
\right]^{q-n}
[\tau(1)]^q [1 + R_r(1)]^{-n},
\end{align}
which has the form of (8.10) with $n$ replaced by $-n$. 
The rest of the calculation is the same as that for (8.10), and we find that (8.11) holds with the replacement $n \to -n$. 
From (8.18a) and the first equation of (8.17), it follows that 
\begin{equation}
L_-^{-n} \mu^m = \frac{(m-n-1)!}{(m-1)!}\, \mu^{m-n}, \qquad
L_+^{-n} \mu^m = (-1)^n \frac{ m!}{(m+n)!}\, \mu^{m+n}.
\end{equation}
Using (8.22) and (7.19), we can see that (8.12) remains valid with $n$ replaced by $-n$. Therefore, (8.13) also holds with $n$ replaced by $-n$, which  verifies (8.16).

Although (8.7), (8.14), and (8.16) are representation-independent expressions, the vectors $\Psi_0$ and $\Phi_0$ satisfying the required conditions  may not necessarily exist in a finite-dimensional representation. 
These general formulas can be fully utilized only in infinite-dimensional representations like the one in Sec.~VII.

\section{Simplified expressions with $\boldsymbol{q=0}$}
Each of the expressions in (5.5), (8.7), and (8.14) contains a denominator which has the form of

\noindent
$\langle \Phi_0, U(\infty, -\infty) L_+^n L_-^n \Psi_0 \rangle$ with some $n$. This denominator does not have a meaning by itself. As in (4.16), we need to let $z' \to -\infty$ and $z \to +\infty$ after dividing the numerator by the denominator. 
It is much convenient if we can get rid of such a denominator.
This can be achieved by using the representation introduced in Sec.~VII. 

In this section, we solely work with the infinite-dimensional representation given by (7.3), so the expressions derived here are not representation independent.

In (7.12), the denominator on the right-hand side has the form of (7.23), which is proportional to $[\tau(z,z')]^q$. This denominator would become a harmless constant if we could choose $q=0$. We cannot directly set $q=0$, however, since $L_- \Psi_{0,q}=0$ for $q=0$ [and, accordingly, (7.23) has a factor which vanishes for $q=0$]. 
We have to consider a limit in which $q$ approaches zero. Since the index $q$ in (7.4) was defined as an integer, it is necessary to extend $q$ to real numbers in order to take the limit $q \to 0$. 
Instead of (7.4), we may consider the representation space spanned by
\begin{equation}
\Psi_{p,q} (\xi,\mu) 
= \xi^p \mu^q, 
\qquad
p=0,1,2,\ldots, \quad q=\alpha, \alpha \pm 1, \alpha \pm 2, \alpha \pm 3\ldots,
\end{equation}
where $\alpha$ is a fixed real number, $0 < \alpha < 1$. 
[We need to take both positive and negative integers for $q - \alpha$ in order to make the space closed under the operation of (7.3).]
The inner product in this space can be defined by rewriting (7.5)  in terms of the Gamma function as
\begin{equation}
\langle \Psi_{p'\!, q'}, \Psi_{p, q} \rangle 
= \delta_{p p'} \delta_{q q'} 
\frac{\Gamma(p + 1)\, \Gamma(q + 1)}{\Gamma(p + q)}.
\end{equation}
Relations (7.6) still hold (see Appendix~F) with this inner product.
The vectors satisfying conditions (5.6) can be chosen as $\Psi_0 = \Phi_0 = \Psi_{0,q}$ with any $q$ such that $q - \alpha$ is an integer. Since we are going to take the limit $q \to 0$, we shall use $q = \alpha$.

Equations~(7.23) and (7.15) are valid even if $q$ is not an integer. With $q = \alpha$, they read 
\begin{equation}
\langle \Psi_{0,\alpha}, U(z,z') L_+ L_- \Psi_{0,\alpha} \rangle 
= - \alpha^2 (\alpha +1) [\tau(z,z')]^\alpha,
\end{equation}
\begin{equation}
(L_- + K_+) U(y,z') \Psi_{0,\alpha} = \alpha \mu \, 
\frac{1 + R_r(y,z')}{1 - \xi R_r(y,z')}
\left[
\frac{\mu \tau(y,z')}{1 - \xi R_r(y,z')}
\right]^\alpha.
\end{equation}
In Eq.~(7.12), we set $q = \alpha$, substitute (9.3) into the denominator, and then take the limit $\alpha \to 0$ before $z' \to -\infty$ and $z \to + \infty$. This yields
\begin{equation}
2 i k G(x,y) = \lim_{\mathstrut z \to + \infty} \lim_{\mathstrut z' \to - \infty}
\lim_{\mathstrut \alpha \to 0}
\frac{-1}{\alpha^2}
\langle \Psi_{0,\alpha}, U(z,x) (L_+ - K_-) U(x,y) (L_- + K_+) U(y,z') \Psi_{0,\alpha} \rangle.
\end{equation}
Since $(L_+ - K_-)^\dagger = - (L_- + K_+)$ [see~(7.6)], we can write the right-hand side of (9.5) as
\begin{equation}
\lim_{\mathstrut z \to + \infty} \lim_{\mathstrut z' \to - \infty}
\lim_{\mathstrut \alpha \to 0}
\frac{1}{\alpha^2}
\langle (L_- + K_+) U^\dagger (z,x) \Psi_{0,\alpha}, U(x,y) (L_- + K_+) U(y,z') \Psi_{0,\alpha} \rangle,
\end{equation}
where the adjoint of the evolution operator $U$ is [see (3.6) and (7.6)]
\begin{equation}
U^\dagger =\left(e^{- R_r J_+} \tau^{2 J_3} e^{R_l J_-} \right)^\dagger
= e^{-R_l^* J_+} (\tau^*)^{2 J_3} e^{R_r^* J_-}.
\end{equation}
We define
\begin{subequations}
\begin{align}
\Lambda_r(y) &\equiv \lim_{z' \to -\infty} \lim_{\alpha \to 0} 
\frac{1}{\alpha}\,(L_- + K_+) U(y,z') \Psi_{0,\alpha},
\\
\Lambda_l(x) &\equiv \lim_{\,z \to +\infty\,} \lim_{\alpha \to 0} 
\frac{1}{\alpha}\,(L_- + K_+) U^\dagger (z,x) \Psi_{0,\alpha}.
\end{align}
\end{subequations}
The right-hand side of (9.8a) is obtained from (9.4). 
By comparing (9.7) with (3.6), we find that $\Lambda_l(x)$ is obtained from $\Lambda_r(y)$ by replacing $R_r(y,-\infty)$ with $R_l^*(\infty,x)$. Thus,　
\begin{equation}
\Lambda_r(y) = \mu \, \frac{1 + R_r(y,-\infty)}{1 - \xi R_r(y,-\infty)}, 
\qquad
\Lambda_l(x) = \mu \, \frac{1 + R_l^*(\infty,x)}{1 - \xi R_l^*(\infty,x)}.
\end{equation}
We can express Eqs.~(9.9) in terms of the basis vectors defined by (7.4) as
\begin{equation}
\Lambda_r = (1 + R_r) \sum_{p=0}^\infty (R_r)^p \,\Psi_{p,1}, \qquad
\Lambda_l = (1 + R_l^*) \sum_{p=0}^\infty (R_l^*)^p \,\Psi_{p,1}.
\end{equation}
From (9.5), (9.6), and (9.8), we obtain the simple expression
\begin{equation}
2 i k G(x,y) = \langle \Lambda_l(x), U(x,y) \Lambda_r(y) \rangle.
\end{equation}
In the same way, from (8.7) we can derive
\begin{equation}
[2 i k G(x,y)]^n
=
\frac{(-1)^{n+1}}{n!\, (n-1)!}\,
\langle \Lambda_l(x),(L_+ - K_-)^{n-1} U(x,y) (L_- + K_+)^{n-1} \Lambda_r(y) \rangle.
\end{equation}
It can be seen from (7.14) that both $\Lambda_r$ and $\Lambda_l$ have the remarkable property
\begin{equation}
(L_- + K_+)^n \Lambda_r(y) = n! \, [\Lambda_r(y)]^{n+1}, \qquad
(L_- + K_+)^n \Lambda_l(x) = n! \, [\Lambda_l(x)]^{n+1}.
\end{equation}
Using (9.13), we can also write (9.12) as
\begin{equation}
[2 i k G(x,y)]^n
= (1/n)
\langle \Lambda_l^n(x), U(x,y) \Lambda_r^n(y) \rangle,
\end{equation}
where $\Lambda_l^n(x) \equiv [\Lambda_l(x)]^n$, \,$\Lambda_r^n(y) \equiv [\Lambda_r(y)]^n$. Similarly, expressions (8.14) can be written as
\begin{subequations}
\begin{align}
& (2 i k)^2 G(x_2, y_2) G(x_1, y_1) 
\nonumber \\*
& \qquad
=
-\frac{1}{2}
\langle \Lambda_l(x_2), U(x_2,x_1) (L_+ - K_-) U(x_1, y_2)
(L_- + K_+) U(y_2, y_1) \Lambda_r(y_1) \rangle,
\\
& (2 i k)^3 G(x_3, y_3) G(x_2, y_2) G(x_1, y_1) 
\nonumber \\*
& \qquad
=
\frac{1}{12}
\langle \Lambda_l(x_3), U(x_3,x_2) (L_+ - K_-) U(x_2,x_1) (L_+ - K_-) U(x_1, y_3)
\nonumber \\*
& \qquad \qquad\qquad \qquad \qquad \qquad
\times
(L_- + K_+) U(y_3, y_2) (L_- + K_+) U(y_2, y_1) \Lambda_r(y_1) \rangle.
\end{align}
\end{subequations}

As a simple example of calculation, let us consider $G(x,y)$ for $y=x$. 
Substituting (9.10) into (9.11) with $y=x$, and using $\langle \Psi_{p'\!,1}, \Psi_{p,1} \rangle = \delta_{p p'}$, we have
\begin{multline}
2 i k G(x,x) =
\langle \Lambda_l(x), \Lambda_r(x) \rangle 
= (1 + R_l) (1 + R_r) \sum_{p=0}^\infty \sum_{p'=0}^\infty
(R_l)^{p'} (R_r)^p \langle \Psi_{p'\!,1}, \Psi_{p,1} \rangle
\\*
 = (1 + R_l) (1 + R_r) \sum_{p=0}^\infty (R_l R_r)^p 
= \frac{[1 + R_l(\infty,x)][1 + R_r(x, -\infty)]}{1 - R_l(\infty,x) R_r(x, -\infty)}.
\end{multline}
This agrees with (7.24) for the case $y=x$ [where $\tau(x,x)=1$ and
$R_r(x,x)=R_l(x,x)=0$]. 

It is also possible to write (9.11), (9.14), and (9.15) using only $\Lambda_r$, without $\Lambda_l$. 
Obviously, $U(L_+ - K_-) U (L_- + K_+) U\Psi_{0, \alpha}$ can be expressed as a linear combination of $\Psi_{p, \alpha}$ with various~$p$.
(Note that $L_- + K_+$ raises and $L_+ - K_-$ lowers the power of $\mu$ by one, whereas $U$ neither raises nor lowers it.)
Since $\langle \Psi_{0,\alpha}, \Psi_{0,\alpha} \rangle = \alpha$, we can see that
$\langle \Psi_{0,\alpha}, \Phi \rangle 
= 
\alpha c_0 = \alpha \mu^{-\alpha}
\Phi 
\, \big\vert_{\xi=0}$ if $\Phi = \sum_{p=0}^\infty c_p \Psi_{p, \alpha}$.
Thus,
\begin{multline}
\langle \Psi_{0,\alpha}, U(3) (L_+ - K_-) U(2) (L_- + K_+) U(1) \Psi_{0,\alpha} \rangle
\\*
\qquad
 = \alpha \mu^{-\alpha} U(3) (L_+ - K_-) U(2) (L_- + K_+) U(1) \Psi_{0,\alpha} \big\vert_{\xi=0}.
\end{multline}
Substituting (9.17) into (9.5), we obtain
\begin{equation}
2ik G(x,y) =
- \,U(\infty,x) (L_+ - K_-) U(x,y) \Lambda_r(y) \big\vert_{\xi=0},
\end{equation}
where $U(\infty,x) \cdots \vert_{\xi=0}$ is shorthand for
$\lim_{z \to \infty} \bigl\{ U(z,x) \cdots \vert_{\xi=0} \bigr\}$.
Similarly, corresponding to (9.14) and (9.15), we have
\begin{equation}
[2ik G(x,y)]^n =
\frac{(-1)^n}{n!}\, U(\infty,x) (L_+ - K_-)^n U(x,y) 
 \Lambda_r^n(y) \big\vert_{\xi=0},
\end{equation}
\begin{subequations}
\begin{align}
& (2ik)^2 G(x_2,y_2) G(x_1,y_1) =
\frac{1}{2}\, U(\infty,x_2) (L_+ - K_-) U(x_2,x_1)(L_+ - K_-)
\nonumber \\*
& \qquad \qquad \qquad \qquad \qquad \qquad \qquad \qquad 
\times
 U(x_1,y_2)(L_- + K_+) U(y_2,y_1) \Lambda_r(y_1)
\big\vert_{\xi=0},
\\*
& (2ik)^3 G(x_3,y_3) G(x_2,y_2) G(x_1,y_1)
\nonumber \\*
& \qquad
= 
- \frac{1}{12}\,
U(\infty,x_2) (L_+ - K_-) U(x_3,x_2) (L_+ - K_-) U(x_2,x_1) (L_+ - K_-) 
\nonumber \\*
& \qquad \qquad \qquad \qquad 
\times
U(x_1,y_3)(L_- + K_+) U(y_3,y_2) (L_- + K_+) U(y_2,y_1)\Lambda_r(y_1)
 \big\vert_{\xi=0}.
\end{align}
\end{subequations}
Such asymmetric expressions are often more convenient in practice than the symmetric expressions that use both $\Lambda_r$ and $\Lambda_l$.

\section{Multiple scattering and the infinite-dimensional representation of $\boldsymbol{SL(3,\mathbf{C})}$}
The Green function $G(x, y)$ can be interpreted as a superposition of waves propagating from point $y$ to point $x$ undergoing multiple scattering by the potential.
In this section, let us explain how the infinite-dimensional representation of $SL(3,\mathbf{C})$ introduced in Sec.~VII gives a natural description of multiple scattering processes.

The multiple-scattering picture becomes simpler when scattering is expressed in terms of the function $f$ rather than $V_\mathrm{S}$. 
As shown in Appendix~J, the Green function multiplied by $2 ik$ can be written as a formal series in $f$,
\begin{align}
2 i k G(x,y) 
&= e^{ik (x-y)} + 
\int_{-\infty}^y dz_1 
\,e^{ik(x - z_1)}
f(z_1) 
\,e^{- ik(z_1 - y)} 
-\int_x^\infty dz_1 
\, e^{- ik(x - z_1)} 
f(z_1) 
\, e^{ik(z_1 - y)}
\notag \\
& \quad
- \int_{-\infty}^y  dz_1 \int_x^\infty dz_2 
\, e^{- ik(x - z_2)} 
f(z_2)
\, e^{ik(z_2 - z_1)} 
f(z_1)
\, e^{- ik(z_1 - y)}
\nonumber \\
& \quad
- \int_y^\infty dz_1 \int_{-\infty}^{\min(x,z_1)}dz_2 
\, e^{ik(x - z_2)}
f(z_2) 
\, e^{- ik(z_2 - z_1)} 
f(z_1)
\, e^{ik(z_1 - y)} 
+ \cdots .
\end{align}
[The terms of general order in this series are shown in (J5) and (J6) of Appendix~J.]
The first term on the right-hand side, $e^{ik(x-y)}$, describes a wave propagating rightwards
\cite{Note5}
 from~$y$ to~$x$.  
The second and third terms can be interpreted as describing a wave that propagates from $y$ to $z_1$, gets scattered at $z_1$, and then propagates to $x$.
In the second term, the factor $e^{- ik(z_1 - y)}$ should be understood as a wave propagating leftwards from $y$ to $z_1$. The wave changes direction at $z_1$ and then goes rightwards to $x$, as can be seen from the factor $e^{ik(x - z_1)}$.
In this manner, each term in (10.1) represents a multiple-scattering event. 
Each single scattering at position~$z$ gives a factor $+f(z)$ if the wave comes to $z$ from the right (as in the second term) and $-f(z)$ if it comes from the left (as in the third term). 
It should be noted that the wave always changes direction when scattered. 
There is no ^^ ^^ forward scattering" if scattering is described in terms of $f$. (This is an advantage of using $f$ instead of $V_\mathrm{S}$.) 
The first five terms in (10.1) can thus be expressed graphically as in Fig.~3.%
%
%        Figure 3    
%
\begin{figure}
\includegraphics[scale=0.7]{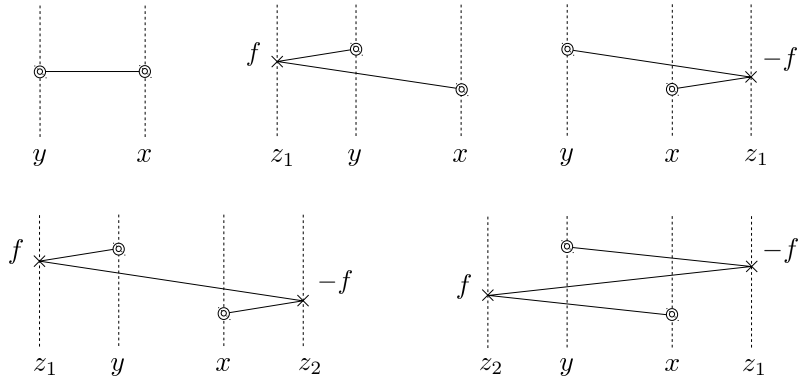}
\caption{
Graphical representation of the first five terms on the right-hand side of (10.1). 
The endpoints at $y$ and $x$ are marked by double circles. Crosses denote the points where scattering occurs. 
In the fifth term, $z_1$ and $z_2$ can move in the region $y \leq z_1$ and $z_2 \leq x$ with the restriction $z_2 \leq z_1$.
}
\end{figure}
The connection between (10.1) and Fig.~3 is given by the rules summarized in  Fig.~4. 

A higher-order term in (10.1) can be expressed graphically as in Fig.~5, corresponding to a process in which a wave starting from $y$ is successively scattered at $z_1, z_2, z_3,\ldots$ before arriving at~$x$.
%
%
%        Figure 4    
%
\begin{figure}
\includegraphics[scale=0.7]{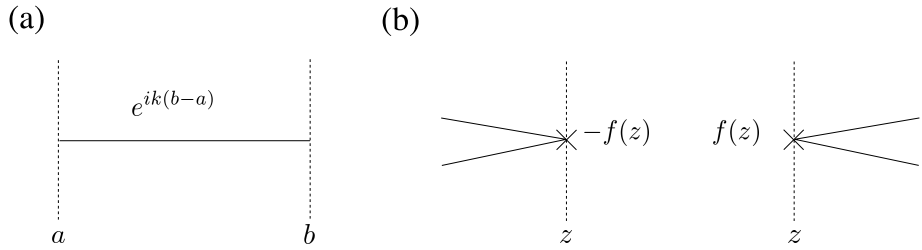}
\caption{
Rules for evaluating the diagram. (a) Each line segment connecting two points $a$ and $b$ ($a < b$) has the value $\exp[ik(b - a)]$. (b) Scattering at point $z$ gives a factor $\pm f(z)$. 
}
\end{figure}
%
%
%      
%
%
%
%        Figure 5    
%
\begin{figure}
\includegraphics[scale=0.7]{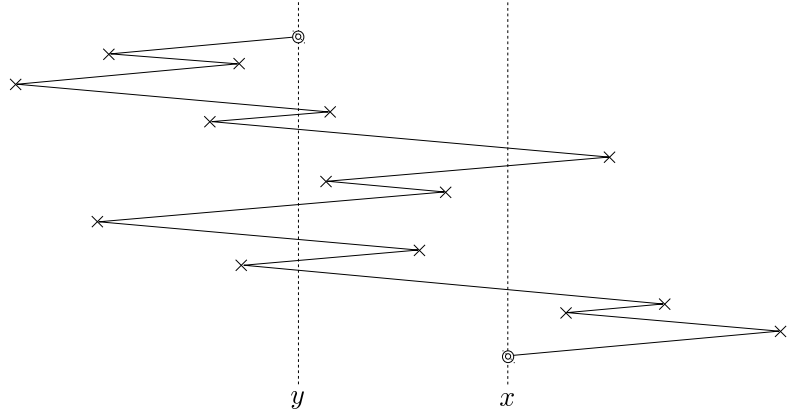}
\caption{
A diagram representing a path that connects two endpoints $y$ and $x$ in one-dimensional space. The horizontal direction of the diagram corresponds to the spatial coordinate. (The vertical direction does not have any particular meaning.) The Green function $G(x,y)$ multiplied by $2ik$ is obtained as the sum of all diagrams like this one. 
}
\end{figure}
Figure~5 can be regarded as a diagram representing a path in one-dimensional space. 
The diagram has a value given by the rules of Fig.~4, and (10.1) means that $2ik G(x,y)$ is the sum of all such diagrams. (Integrations over $z_1, z_2, z_3 \ldots$ are implied in the sum.) The sum of diagrams can also be thought of as the sum over all paths connecting $y$ and $x$. (We use the word ^^ ^^ path" to mean something like Fig.~5, which corresponds to each term in the multiple-scattering series. It is not a path in the sense of Feynman path integral.)

Let $x_1<x_2$, and suppose that $f(x)=0$ for $x \leq x_1$ and $x \geq x_2$.
Setting $x = x_2$, $y = x_1$ in (7.24) and substituting $R_r(x_1, -\infty) = R_l(\infty, x_2) = 0$ yields $2ik G(x_2, x_1) = \tau(x_2,x_1)$. In other words, the transmission coefficient $\tau(x_2, x_1)$ is equal to $2ik G(x_2, x_1)$ assuming that scattering occurs only within the interval $(x_1,x_2)$. 
Since $ik G(x_2, x_1)$ is the sum over all paths connecting $x_1$ and $x_2$, we can see that $\tau(x_2,x_1)$ is the sum over all paths that start from $x_1$ and end at $x_2$ without leaving the interval $(x_1,x_2)$ [Fig.~6(a)]. 
%
%
%        Figure 6      
%
\begin{figure}
\includegraphics[scale=0.7]{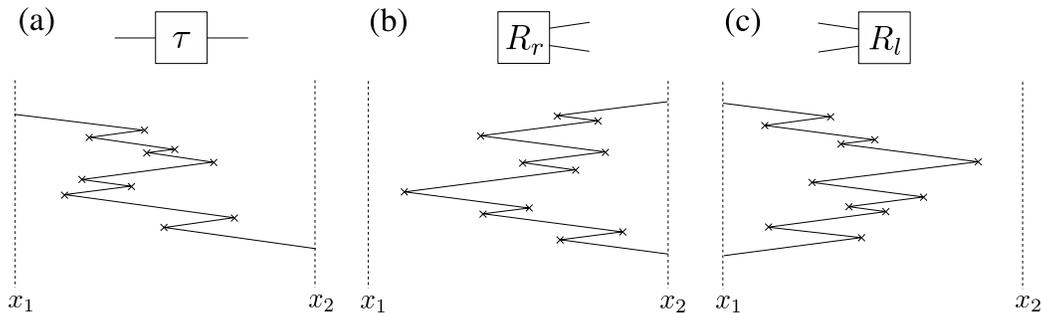}
\caption{
Diagrams for (a) the transmission coefficient $\tau(x_2,x_1)$, (b) the right reflection coefficient $R_r(x_2,x_1)$, and (c) the left reflection coefficient $R_l(x_2,x_1)$.
}
\end{figure}
Similarly, setting $x = y = x_2$ in (7.24) and using $R_r(x_2,-\infty) = R_r(x_2, x_1)$ with $R_l(\infty, x_2) = 0$ and $\tau(x_2,x_2) = 1$, we find that $2ikG(x_2,x_2) = 1 + R_r(x_2,x_1)$ if scattering occurs only within the interval $(x_1,x_2)$. 
Therefore, the right reflection coefficient $R_r(x_2, x_1)$ is the sum over all paths that start from and return to $x_2$ without leaving the interval $(x_1,x_2)$ [Fig.~6(b)]. In the same way, the left-reflection coefficient $R_l(x_2,x_1)$ is the sum over all paths starting from and returning to $x_1$, confined within the interval $(x_1,x_2)$. [Fig~6(c)].

Now let us see how the formalism presented in Sec.~VII works as a method for taking the sum over such paths. 
%
%
%        Figure 7      
%
\begin{figure}
\includegraphics[scale=0.7]{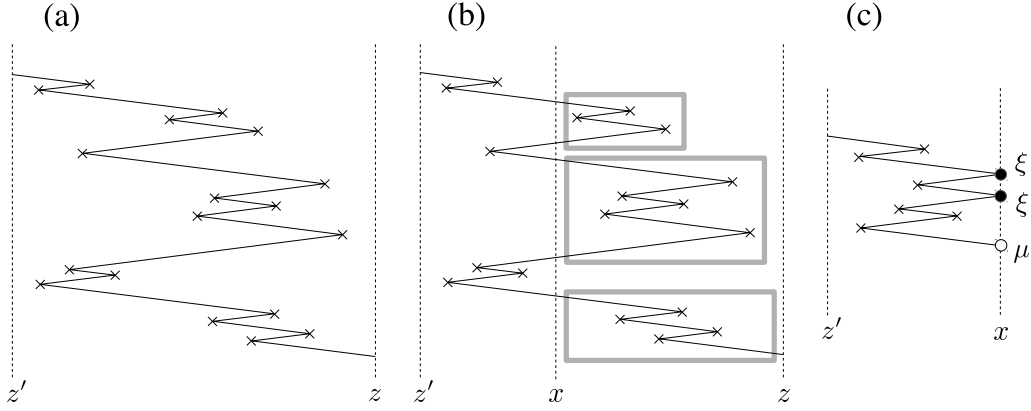}
\caption{
(a) A diagram representing a path in the interval $(z',z)$. 
(b) Parts of the path in the interval $(x, z)$ (rectangular boxes) are contracted to black and white circles.
(c) The diagram thus obtained from (b).
}
\end{figure}
Figure~7(a) shows a diagram representing a path that runs through the interval $(z',z)$. 
The sum of all such diagrams is the transmission coefficient $\tau(z, z')$ [see Fig.~6(a)].
Let us take an arbitrary point $x$ between $z'$ and $z$. We truncate this diagram at $x$, retaining only the part to the left of $x$. 
Each fragment of the path to the right of $x$ [rectangular boxes in Fig.~7(b)] is contracted to a black circle if it starts from $x$ and returns to $x$. It is contracted to a white circle if it connects the two points $x$ and $z$. 
In this way, we obtain a reduced diagram with white and black circles as in Fig.~7(c). We assign a $\xi$ to a black circle and a $\mu$ to a white circle. Then, such a diagram takes a value in the vector space spanned by the vectors $\Psi_{p, q}$ of (7.4). For instance, the value of the  diagram in Fig.~7(c) is $C\Psi_{1,2} = C\mu \xi^2$, where the scalar factor $C$ is the value of the path in the interval $(z',x)$  calculated according to the rules of Fig.~4.

The general equation satisfied by the evolution operator $U$ is Eq.~(3.1). 
In the square brackets on the right-hand side of this equation, the first term $2ik J_3$ describes free evolution and the second term $-2fJ_1$ describes scattering.
In the representation given by (7.3), the scattering term reads $-2 f J_1 = - f(\partial/\partial \xi) + f \xi^2 (\partial/\partial \xi) + f\xi \mu (\partial/\partial \mu)$. Hence, scattering is described by the three operators $- f(\partial/\partial \xi)$, $f \xi^2 (\partial/\partial \xi)$, and $f\xi \mu (\partial/\partial \mu)$. The graphical meaning of these differential operators is shown in Fig.~8.
%
%
%        Figure 8      
%
\begin{figure}
\includegraphics[scale=0.7]{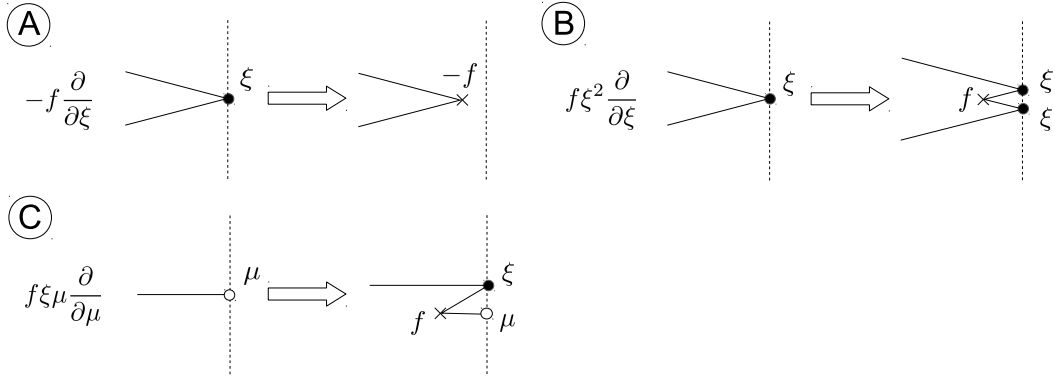}
\caption{
The meaning of each term in $-2fJ_1 = -f(\partial/\partial \xi) + f \xi^2 (\partial/\partial \xi) + f \xi \mu (\partial/\partial \mu)$.
The operator $\partial/\partial \xi$ annihilates a black circle, $\xi^2(\partial/\partial \xi)$ turns a black circle into two black circles, and $\xi \mu (\partial/\partial \mu)$ turns a white circle into a pair of white and black circles. The factor $\pm f$ corresponds to a scattering vertex as in Fig.~4(b).
}
\end{figure}
Paths with turning points are created by the action of these three operators.
For example, the path in Fig.~9(a) is generated as shown in Fig.~9(b).
%
%
%        Figure 9      
%
\begin{figure}
\includegraphics[scale=0.7]{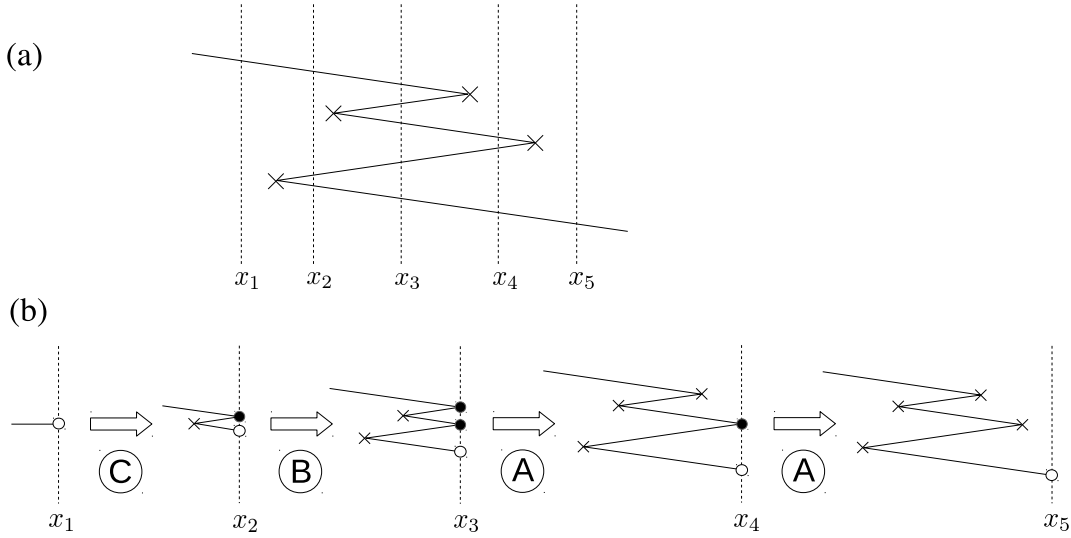}
\caption{
The path shown in (a) is generated as in (b), by a combination of the processes A -- C of Fig.~8.
}
\end{figure}
Meanwhile, the operator $2 J_3=2 \xi (\partial/\partial \xi) + \mu (\partial/\partial \mu)$ counts the number of lines at each cross section of the diagram. 
In the example of Fig.~9(a), the operator $2J_3$ gives the value $m=1$, $3$, $5$, and $3$, $1$ at $x_1$, $x_2$, $x_3$, $x_4$, and $x_5$, respectively. Therefore, the free evolution term $2ik J_3$ gives rise to the factor $\exp[mik(b-a)]$ between two points $a$ and $b$ where there are $m$ lines. This is the value of the line segments given by the rule of Fig.~4(a).

The operator $U$ describes the evolution of the path from left to right. 
As in Fig.~9(b), the evolution of a single path begins with a white circle, which corresponds to $\Psi_{0,1} = \mu$. 
The action of the operator $U(z, z')$ on $\Psi_{0,1}$ generates diagrams like the one in Fig.~10(a), which may include an arbitrary number of black circles. (There is only one white circle.) 
%
%
%        Figure 10      
%
\begin{figure}
\includegraphics[scale=0.7]{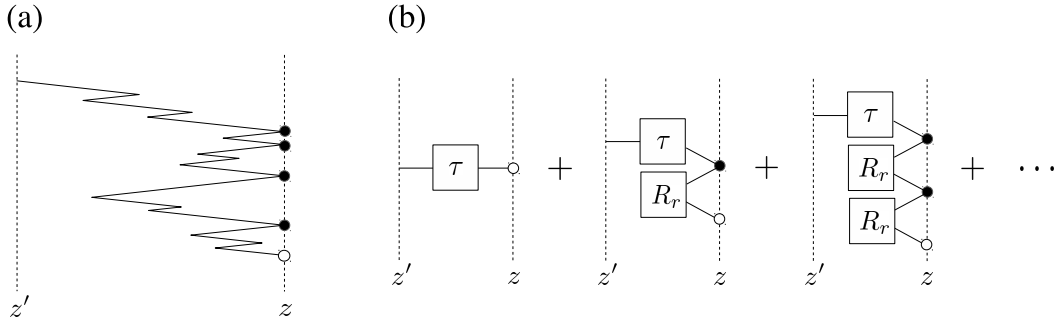}
\caption{
(a) A diagram contributing to $U(z,z') \Psi_{0,1}$. 
(b) Expression of $U(z,z') \Psi_{0,1}$ in terms of $\tau$ and $R_r$.
}
\end{figure}
The vector $U(z, z') \Psi_{0,1}$ is the sum of all such diagrams. 
We can write $U(z, z') \Psi_{0,1} = \sum_{m=0}^\infty  C_m \mu \,\xi^m$, where $C_m \mu \,\xi^m$ denotes the sum of all diagrams with one white circle and $m$ black circles. From the graphical interpretation of the transmission and reflection coefficients given by Figs.~6(a) and 6(b), we can see that $C_m = \tau(z, z') [R_r(z, z')]^m$.
Therefore, 
\begin{equation}
U(z, z') \Psi_{0,1} =
 \mu \tau +  \mu \tau \xi R_r + \mu \tau (\xi R_r)^2 + \mu \tau (\xi R_r)^3 +\cdots =\frac{\mu \tau}{1 - \xi R_r} = \hat T(z, z').
\end{equation}
[The last equality is definition (7.10).] Thus, $U(z, z') \mu = \hat T(z, z')$ [see (7.9)]. The geometric series in (10.2) can be expressed graphically as in Fig.~10(b).

Taking the inner product of $U(z, z') \Psi_{0,1}$ with $\Psi_{0,1}$ picks out $C_1$ from $U \Psi_{0,1} = \sum_{m=0}^\infty  C_m \mu \,\xi^m$. 
That is, $\langle\Psi_{0,1}, U(z, z') \Psi_{0,1}\rangle =  C_1 = \tau(z, z')$. Graphically, taking the inner product with $\Psi_{0,1}$ means selecting the diagrams with no black circles.  
We retrieve Fig.~7(a) from Fig.~10(a) by picking out only the diagrams with no black circles (and then removing the white circle). 

Let us assume that $z' < y \leq x < z$. (We will let $z \to +\infty$ and $z' \to -\infty$ later.) 
In order to consider the Green function $G(x, y)$, we need to deal with a path that has endpoints at $y$ and $x$. The endpoints of a path are created by the operators $K_+ + L_- = \mu^2 (\partial/\partial \mu) +  \mu (\partial/\partial \xi) + \xi \mu (\partial/\partial \xi)$ and $K_- - L_+= (\partial/\partial \mu) + \xi (\partial/\partial \mu)$. The meaning of each term in these operators is illustrated in Fig.~11. 
%
%
%        Figure 11      
%
\begin{figure}
\includegraphics[scale=0.7]{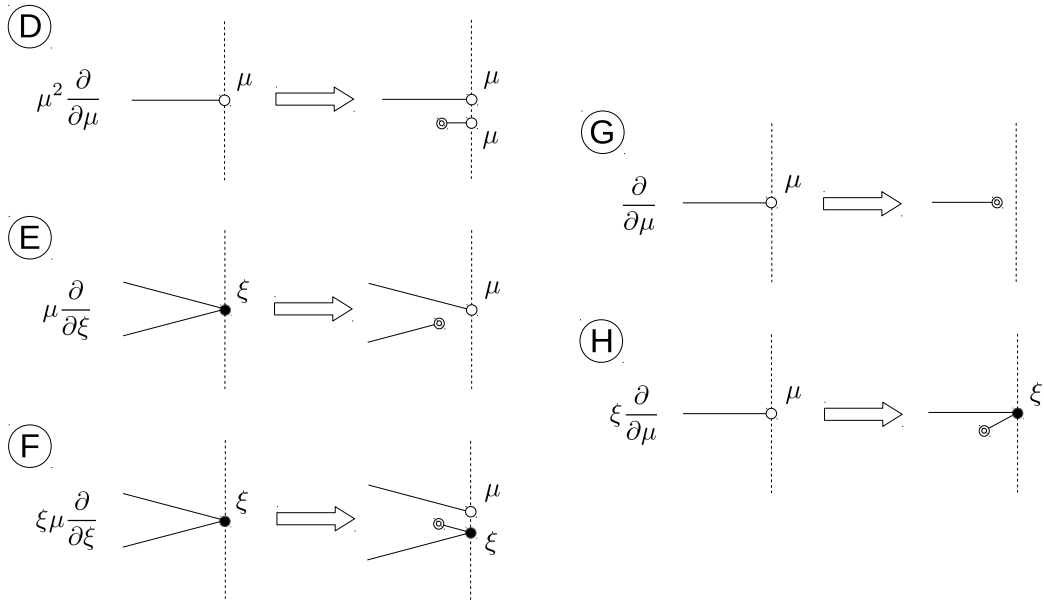}
\caption{
The meaning of each term in $K_+ +  L_- = \mu^2 (\partial/\partial \mu) + \mu (\partial/\partial \xi) + \xi \mu (\partial/\partial \xi)$ and $ K_- -L_+= (\partial/\partial \mu) + \xi (\partial/\partial \mu)$.
As in Fig.~3, a double circle means an endpoint of a path.
}
\end{figure}
Each of the three terms in $K_+ + L_-$ increases the number~$n$ of white circles by one, and each of the two terms in $K_- - L_+$ decreases $n$ by one.
Figure~12(a) shows three examples of a path with endpoints at $y$ and $x$. Diagrams including such paths can be generated by the action of the operators $U$, $K_+ + L_-$, and $K_- - L_+$ on the vector $\Psi_{0,1}$ as shown in Fig.~12(b).
%      
%
%        Figure 12
%
\begin{figure}
\includegraphics[scale=0.7]{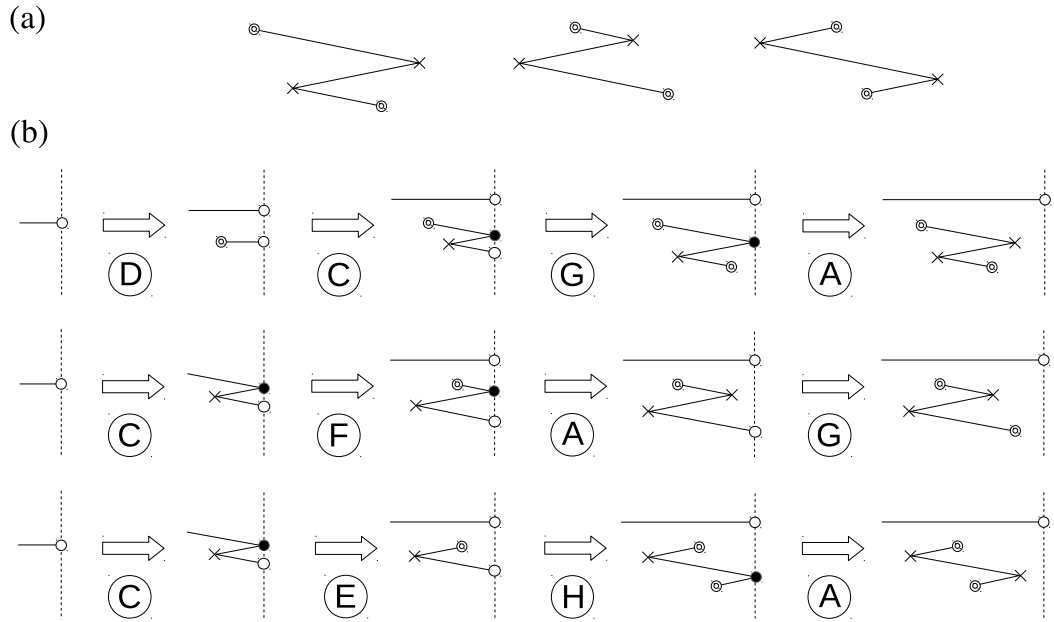}
\caption{
The three paths in (a) are generated through the processes shown in (b), by the action of the operators $U$ (A -- C of Fig.~8), $K_+ + L_-$ (D -- F of Fig.~11), and $K_- + L_+$ (G and H of Fig.~11) on the vector $\Psi_{0,1}$.
}
\end{figure}
More generally, the successive action of the operators $U(y,z')$, $K_+ + L_-$, $U(x,y)$, $K_- - L_+$, and $U(z,x)$ on $\Psi_{0,1}$ generates diagrams like Fig.~13(a), with an arbitrary number of black circles.
%      
%
%        Figure 13
%
\begin{figure}
\includegraphics[scale=0.7]{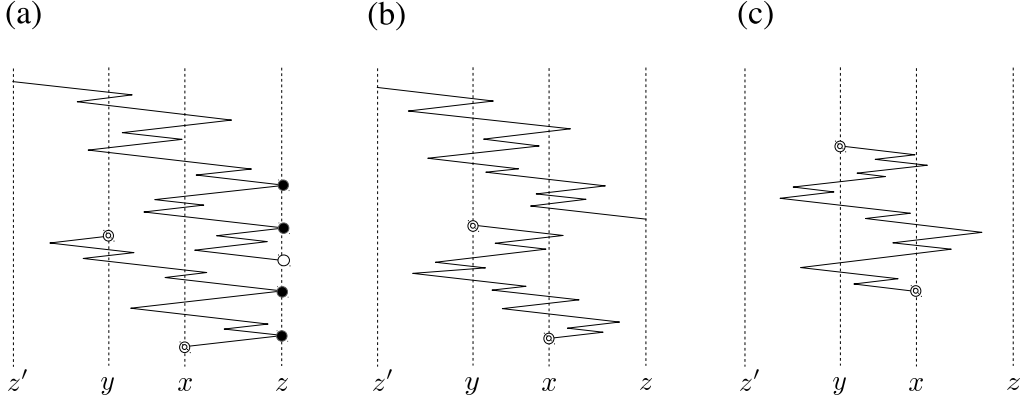}
\caption{
(a) A typical diagram contributing to (10.3). (b)  A typical diagram contributing to (10.4).
(c) A diagram representing a single path restricted within the interval $(z', z)$.
The sum of~(b) is equal to the sum of~(c) multiplied by $\tau(z,z')$.
}
\end{figure}
The vector 
\begin{equation}
U(z,x) (K_- - L_+) U(x,y) (K_+ + L_-) U(y,z') \Psi_{0,1}
\end{equation}
is the sum of all such diagrams. Taking the inner product of this vector with $\Psi_{0,1}$ singles out diagrams with no black circles. Hence, the quantity
\begin{equation}
\langle\Psi_{0,1}, U(z,x) (K_- - L_+) U(x,y) (K_+ + L_-) U(y,z') \Psi_{0,1}\rangle
\end{equation}
is the sum of all diagrams such as the one in Fig.~13(b).

The diagram in Fig.~13(b) [as well as the ones in Fig.~13(a) and Fig.~12(b)] consists of two connected parts.
Along with the path connecting the endpoints $y$ and $x$, it includes another path running through the whole interval $(z, z')$.
This auxiliary path is a diagram for the transmission coefficient as in Fig.~6(a), and the sum of all such diagrams amounts to $\tau(z,z')$. Therefore, (10.4) divided by $\tau(z, z')$ is the sum of the diagrams like the one in Fig.~13(c). 
Unlike Fig.~4, the diagram in Fig.~13(c) is restricted within the interval $(z',z)$. 
Removing this restriction by letting $z' \to -\infty$ and $z \to +\infty$, we obtain $2ik G(x, y)$. Thus,
\begin{equation}
\lim_{z \to \infty} \lim_{z' \to -\infty} 
\frac{\langle\Psi_{0,1}, U(z, x) (K_- - L_+) U(x, y) (K_+ + L_-)U(y, z') \Psi_{0,1}\rangle}{2\,\tau(z, z')}
= 2ik G(x, y).
\end{equation}
(The origin of the factor 2 in the denominator on the left-hand side is explained in Fig.~14.)
%      
%
%        Figure 14
%
\begin{figure}
\includegraphics[scale=0.8]{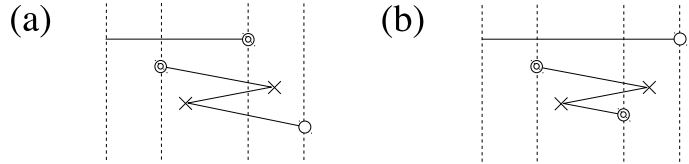}
\caption{
Diagrams (a) and (b) have the same value, so we can replace (a) with (b).
Thus, diagrams of type~(a) (in which the two endpoints are disconnected) can be disregarded if diagrams of type~(b) (with the endpoints connected) are counted twice. This gives the factor 2 in (10.5).
}
\end{figure}
This is the graphical meaning of Eq.~(7.12) with $q=1$. 

The same interpretation is possible for any positive integer $q = n$. 
Each diagram contributing to $\langle\Psi_{0,n}, U (K_- - L_+) U (K_+ + L_-)U \Psi_{0,n}\rangle$ consists of $n+1$ connected parts, namely, one path like Fig.~13(c) and $n$ auxiliary paths penetrating the interval $(z',z)$. To obtain $2ik G(x, y)$, we need to divide out the contribution from the $n$ auxiliary paths, which is equal to $[\tau(z,z')]^n$ multiplied by an appropriate combinatorial factor. This is the meaning of the denominator in (7.12) with $q=n$. 
 
It is also possible to study the evolution of the main path itself, without auxiliary paths. 
To do so, we consider the evolution starting from the point $y$. 
Let us truncate a diagram for $2 i k G(x,y)$ (such as the one in Fig.~5) at $y + \epsilon$ with infinitesimal $\epsilon > 0$. 
Contracting the parts to the right of $y + \epsilon$ as we did in Fig.~7, we obtain a diagram with one white circle and an arbitrary number of black circles. 
Such diagrams are classified into two types as shown in Fig.~15(a), according to the initial direction of the path at $y$ (rightward in the upper diagram, leftward in the lower one). 
%      
%
%        Figure 15
%
\begin{figure}
\includegraphics[scale=0.7]{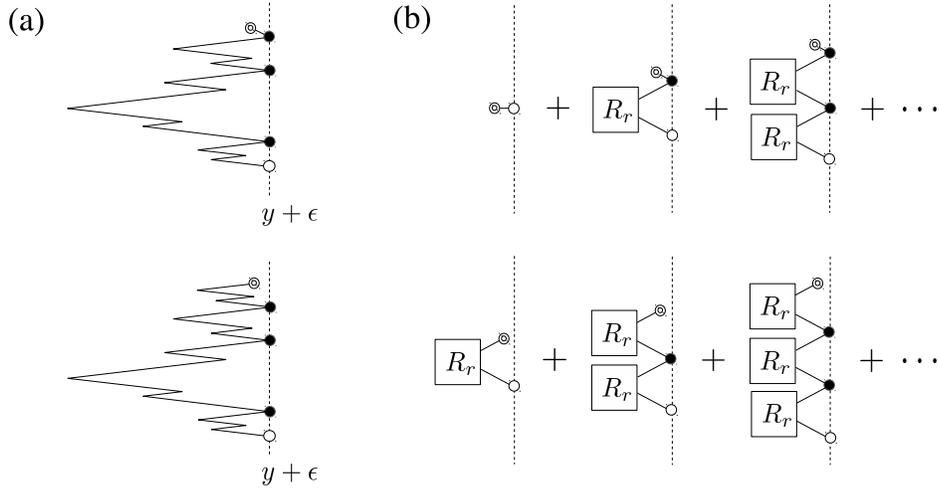}
\caption{
(a) Two types of diagrams obtained by truncating a single-line path at $y + \epsilon$.  
(b) For each type in (a), the sum of diagrams can be expressed in terms of $R_r$, in the same way as in Fig.~10(b).
}
\end{figure}
For each type of Fig.~15(a), the sum of diagrams can be expressed as a geometric series in terms of the reflection coefficient $R_r$, as illustrated in Fig.~15(b).  Putting the two sums in Fig.~15(b) together yields
\begin{equation}
(1 + R_r)[\mu + \mu \xi R_r + \mu (\xi R_r)^2 + \mu (\xi R_r)^3 + \cdots]
= \mu\, \frac{1+R_r}{1 - \xi R_r}.
\end{equation}
This is the vector $\Lambda_r(y)$ defined in (9.9).
The evolution of the single-line path connecting $y$ and $x$ can be studied by examining the action of $U$ and $K_- + L_+$ on $\Lambda_r(y)$. This is what was done in Sec.~IX.

So far, our discussion has been formal and we have not paid attention to the convergence of the various sums of diagrams. 
Actually, these sums are not necessarily convergent. 
To begin with, the expression of the Green function as a perturbation series [Eq.~(10.1) or Eq.~(J1)] is, in general, divergent. The sums of diagrams for the transmission and reflection coefficients in Fig.~6 are not necessarily convergent, either. 
However, such divergence does not cause any problem to our method. 
As can be seen from the calculation in Sec.~VII, our results can always be expressed in terms of transmission and reflection coefficients. Even if the sums of diagrams in Fig.~6 are not convergent, the transmission and reflection coefficients defined by (2.7) are finite and well-defined quantities. 
A divergent series in terms of the bare scattering (denoted by a cross in Fig.~3) can be rearranged into a convergent series in terms of transmission and reflection coefficients. 
For example, the sum of diagrams in Fig.~15(a) may be divergent.
In Fig.~15(b), this possibly divergent multiple scattering series is reinterpreted as a series of multiple reflections. This multiple-reflection series [Eq.~(10.6)] is convergent for $\vert \xi \vert < 1$ since $\vert R_r \vert \leq 1$. 
In (7.24), the Green function is expressed in terms of transmission and reflection coefficients. This expression can be interpreted as a superposition of waves undergoing multiple reflections and transmissions. 
In the calculation leading to (7.24), we are, in effect, adding up multiple reflections, in the same way as in Fig.~10(b) and Fig.~15(b). 
The multiple-scattering picture is thus useful even if the potential is not small enough to be treated as a perturbation. The methods developed in Secs.~VII -- IX can be applied even to potentials that diverge as $x \to \pm \infty$. [In Appendix~J we assumed $f(\pm \infty) =0$, but this assumption is actually not necessary.]

The sum-over-paths view is essential for the understanding of the Schr\"odinger equation, and, as explained above, the representation of $SL(3,\mathbf{C})$ given by (7.3) provides a faithful description of paths in one dimension. 
A cross section of a path corresponds to a vector in the space spanned by (7.4). 
The operator $U$ creates a path in a larger interval by connecting fragments of paths in smaller intervals and thus dictates the evolution of the path from left to right. This description applies not only to a single path but also to a multiple of paths, with endpoints at various positions. Expressions like (8.7) or (8.14) involving products of Green functions can be understood from this perspective.
We can study the evolution of paths and calculate sums over paths by simply considering the action of differential operators on polynomials in $\xi$ and $\mu$, without need of complicated combinatorial arguments. 
Thus, our method provides a simple way for dealing with paths, which can possibly be applied to a wide variety of one-dimensional problems.

\section{Summary and remarks}
In this paper, we studied the structure of solutions of the Schr\"odinger equation described in terms of the Lie group $SL(3,\mathbf{C})$ and its Lie algebra $sl(3,\mathbf{C})$. The key to the algebraic construction of solutions is Eqs.~(3.11), which can be interpreted as representing the wave transmission and reflection shown in Fig.~1. 
Equations (3.11) come from the commutation relations (3.7), and the  $Q_+$ and $Q_-$ satisfying (3.7) correspond to the right-going wave and the left-going wave, respectively. 
In the Lie algebra $sl(3,\mathbf{C})$ defined by (5.1), there are two pairs of such $Q_\pm$, namely, $Q_+ = L_-,\,Q_- = K_+$ and $Q_+ = - K_-, \, Q_- = L_+$. 
Using these operators, the Green function $G(x,y)$ can be expressed as in (5.5), with  different pairs of $Q_\pm$ assigned to the two endpoints $x$ and $y$.
We can also derive more general formulas for products of Green functions, as shown in Sec.~VIII. The existence of two pairs of $Q_\pm$ (each satisfying $[Q_+, Q_-]=0$) is crucial for these general formulas. Expressions like (8.7), (8.14), and (8.16) cannot be constructed with only one pair of $Q_\pm$. 
 
The expressions derived in Secs.~V and VIII are representation independent, that is, they can be used with any representation of the Lie group $SL(3,\mathbf{C})$ and the Lie algebra $sl(3,\mathbf{C})$. 
However, the real merits of this formalism can be appreciated when it is used in combination with an infinite-dimensional representation.  The representation introduced in Sec.~VII is of particular importance. As explained in Sec.~X, this representation faithfully describes the transmission and reflection of waves and provides a natural picture of solutions of the Schr\"odinger equation as waves undergoing multiple scattering processes. 
In this representation, we can deduce the simple formulas of Sec.~IX. 

The formulas obtained in Sec.~IX can be extremely useful for the analysis of the Green function. In Ref.~\onlinecite{formulation}, we used the infinite-dimensional representation of $SL(2,\mathbf{C})$ given by (7.1) for deriving low- and high-energy expansions of reflection coefficients. The same method can be used for $SL(3,\mathbf{C})$ with the representation given by (7.3). 
The function $\Lambda_r$ defined in Sec.~IX is closely related to the generalized reflection coefficient
\cite{Note6} 
$\hat R \equiv \mu^2 R_r/(1 - \xi R_r)$, which was fully studied in Ref.~\onlinecite{formulation}. Using the method of Ref.~\onlinecite{formulation} together with the results of Sec.~IX,  we can systematically study the analytic properties of the Green function and other quantities involving Green functions. This will be discussed elsewhere.

\appendix
\section{PROOF OF (3.6)}
From (1.6) and (2.7), we can derive the differential equations for $\tau$, $R_r$, and $R_l$, 
\begin{gather}
\frac{\partial}{\partial x} R_r(x,y)=2 i k R_r(x,y) + f(x) \{1 -[R_r(x,y)]^2 \}, 
\nonumber \\
\frac{\partial}{\partial x} \tau(x,y) = ik \tau(x,y) - f(x) \tau(x,y) R_r(x,y), 
\qquad
\frac{\partial}{\partial x} R_l(x,y) = -f(x) [\tau(x,y)]^2.
\tag{A1}
\end{gather}
Let us show that the $U(x,y)$ given by (3.6) satisfies (3.1).
Using (A1), we have
\begin{multline}
\frac{\partial}{\partial x} U(x,y) =
[-2ik R_r - f(1 - R_r^2)]  J_+ e^{-R_r J_+} \tau^{2 J_3} e^{R_l J_-} 
\\*
\quad +(2ik - 2 f R_r) e^{-R_r J_+} J_3 \tau^{2 J_3} e^{R_l J_-} 
-f \tau^2 e^{-R_r J_+} \tau^{2 J_3} J_- e^{R_l J_-}.
\tag{A2}
\end{multline}
From the commutation relations (3.5), it easily follows that
\begin{align}
&e^{-R_r J_+} J_3 = (J_3 + R_r J_+) e^{-R_r J_+}, \qquad
\tau^{2 J_3} J_- = \tau^{-2} J_- \tau^{2 J_3}, \nonumber \\
&e^{-R_r J_+} J_- = (J_- - 2 R_r J_3 - R_r^2 J_+) e^{-R_r J_+}.
\tag{A3}
\end{align}
(Note that $[J_+^n, J_-] = 2 n J_3 J_+^{n-1} - n (n-1) J_+^{n-1}$.) 
Therefore, (A2) becomes
\begin{equation}
\frac{\partial}{\partial x} U(x,y) =
(2ik J_3 - f J_+ - f J_-) e^{-R_r J_+} \tau^{2 J_3} e^{R_l J_-} 
= [2ik J_3 - 2 f(x) J_1] U(x,y).
\tag{A4}
\end{equation}
Thus, expression (3.6) indeed satisfies (3.1). 
Since $\tau(y,y)=1$ and $R_r(y,y)=R_l(y,y)=0$, it is obvious that (3.2) is also satisfied.

\section{BEHAVIOR OF $\boldsymbol{\phi_\pm(x)}$ AS $\boldsymbol{x \to \pm \infty}$}
Using (1.8) and (2.1), we can write $\tau(x_0,z') \psi_-(x,z')$ as
\begin{align}
&\tau(x_0,z') \psi_-(x,z')
= 
\frac{\alpha(x,z';k) + \beta(x,z';k)}
{\alpha(x_0,z';k)}
\nonumber
\\*
&\qquad \qquad = 
\frac{\alpha(x,z';k) + \beta(x,z';k)}
{\alpha(x_0,x;k) \alpha(x.z';k) + \beta(x_0,x;-k) \beta(x,z';k)}
= \frac{[1 + R_r(x,z')]\,\tau(x_0,x)}{1 - R_l(x_0,x) R_r(x,z')}.
\tag{B1}
\end{align}
Similarly,
\begin{equation}
\tau(z,x_0) \psi_+(x,z) 
= \frac{[1 + R_l(z,x)]\, \tau(x,x_0)}{1 - R_l(z,x) R_r(x,x_0)}.
\tag{B2}
\end{equation}
The reflection coefficients $R_l(z,x)$ and $R_r(x,z')$ have definite limits\cite{Note2} as $z \to \infty$ and $z' \to -\infty$, respectively.
We can let $z \to \infty$ in (B1) and $z' \to -\infty$ in (B2) to obtain
\begin{equation}
\phi_-(x) = \frac{[1 + R_r(x,-\infty)]\, \tau(x_0,x)}{1 - R_l(x_0,x) R_r(x,-\infty)},
\qquad
\phi_+(x) = \frac{[1 + R_l(\infty,x)]\,\tau(x,x_0)}{1 - R_l(\infty,x) R_r(x,x_0)}.
\tag{B3}
\end{equation}
As shown in Appendix~C, the reflection coefficients always satisfy
\begin{equation}
\vert R_r(x_2,x_1) \vert \leq 1, \qquad
\vert R_l(x_2,x_1) \vert \leq 1,
\tag{B4}
\end{equation}
for any $x_1$ and $x_2$ (either finite or infinite) and $\mathrm{Im}\,k \geq 0$.

Let us study the behavior of $\phi_-(x)$ as $x \to -\infty$.
It can be shown\cite{low} that, as $x \to -\infty$,
\begin{align}
&R_r(x,-\infty) \to 1, \qquad R_l(x_0,x) \to -1 \qquad  
\mbox{if} \ \   f(-\infty) = +\infty, 
\nonumber \\
&R_r(x,-\infty) \to -1, \qquad R_l(x_0,x) \to 1 \qquad 
\mbox{if} \ \   f(-\infty) = -\infty, 
\nonumber \\
&R_r(x,-\infty) \to \frac{i k + \sqrt{c^2 - k^2}}{c}, 
\qquad R_l(x_0,x) \to -\frac{i k + \sqrt{c^2 - k^2}}{c}
\qquad 
\mbox{if} \ \   f(-\infty) = c \neq 0, 
\nonumber \\
&R_r(x,-\infty) \to 0
\qquad
\mbox{if} \ \   f(-\infty) = 0.
\tag{B5}
\end{align}
From (B4) and (B5) we see that, in all cases,
\begin{equation}
\lim_{x \to -\infty} R_l(x_0,x) R_r(x,-\infty) \neq 1,
\qquad
\lim_{x \to -\infty} 
\left\vert
\frac{1 + R_r(x,-\infty) }{1 - R_l(x_0,x) R_r(x,-\infty)}
\right\vert
< \infty.
\tag{B6}
\end{equation}
As $x \to -\infty$, the transmission coefficient $\tau(x_0,x)$ has the asymptotic form
\cite{Note7}
\begin{alignat}{3}
&\tau(x_0,x) = C_1 \exp\left[- \int_x^0 f(z) \,dz + \eta_1(x) \right][1 + o(1)]
& \qquad
& \mbox{if} \ \   f(-\infty) = +\infty, 
\nonumber \\
&\tau(x_0,x) = C_2 \exp\left[\int_x^0 f(z) \,dz + \eta_2(x) \right][1 + o(1)]
& \qquad
& \mbox{if} \ \   f(-\infty) = -\infty, 
\nonumber \\
&\tau(x_0,x) = C_3 \exp\left[\sqrt{c^2 - k^2} \, x +  i \eta_3(x) \right][1 + o(1)]
& \qquad
& \mbox{if} \ \   f(-\infty) = c \neq 0, 
\nonumber \\
&\tau(x_0,x) = C_4 \exp\left[- i k x +  i \eta_4(x) \right][1 + o(1)]
& \qquad
& \mbox{if} \ \  f(-\infty) = 0,
\tag{B7} 
\end{alignat}
where $\eta_i(x) = o(\vert x \vert)$ as $x \to -\infty$ ($i=1,2,3,4$) and $\eta_4$ is real valued if $\mathrm{Im}\,k =0$. 
The factors $C_i$ are independent of $x$. 
In the third line of (B7), the branch of the square root is chosen as $\mathrm{Re}\,\sqrt{c^2 - k^2}>0$ for $\mathrm{Im}\,k >0$. 
We can see from (B7) that $\tau(x_0,x) \to 0$ for $\mathrm{Im}\,k \geq 0$ if $f(-\infty) = \pm \infty$ and $\tau(x_0,x) \to 0$ for $\mathrm{Im}\,k > 0$ if $f(-\infty)=c$ or $0$.
From this and (B6), we find that $\phi_-(x) \to 0$ as $x \to -\infty$ for $\mathrm{Im}\,k >0$ in all cases. In particular, $\phi_-(x) \to 0$ even for real $k$ if $f(-\infty)=\pm \infty$.

In the same way, we can show that $[1 + R_l(\infty,x)]/[1 - R_l(\infty,x) R_r(x,x_0)]$ remains finite as $x \to \infty$ and that $\tau(x,x_0) \to 0$ for $\mathrm{Im}\,k >0$. 
Hence, $\phi_+(x) \to 0$ as $x \to \infty$ for $\mathrm{Im}\,k > 0$. 
In particular, $\phi_+(x) \to 0$ for $\mathrm{Im}\,k \geq 0$ if $f(\infty) = \pm \infty$.

\section{PROOF OF (B4)}
From the first equation of (A1) and its complex conjugate, we have
\begin{equation}
\frac{\partial}{\partial x} \vert R_r (x,y) \vert^2 
=
-4 \,(\mathrm{Im}\,k)\vert R_r \vert^2
+ 2 f(x) (\mathrm{Re}\,R_r)(1 - \vert R_r \vert^2).
\tag{C1}
\end{equation}
For $\mathrm{Im}\,k \geq 0$, we find from (C1) that $(\partial/\partial x) \vert R_r \vert^2 \leq 0$ if $\vert R_r \vert^2=1$. Therefore, $\vert R_r \vert$ cannot become larger than $1$ by crossing $\vert R_r \vert =1$. 
As $\vert R_r(y,y) \vert =0$ for finite $y$, it follows that $\vert R_r(x,y) \vert \leq 1$ for any finite $y$ and $x$. Since $R_r(x,-\infty)$ is the limit $y \to -\infty$ of $R_r(x,y)$, we also have $\vert R_r(x, -\infty) \vert \leq 1$. 
The inequality $\vert R_l \vert \leq 1$ can be proved in the same way.

\section{LESS RESTRICTIVE CONDITIONS ON $\boldsymbol{\Psi}$ AND $\boldsymbol{\Phi}$ IN (3.26)}

In this appendix, we prove the following statements:
(i) If either $\mathrm{Im}\,k>0$ or $f(+\infty) = \pm \infty$, the condition $Q_+^\dagger \Phi = 0$ in (3.26a) can be replaced by $Q_+^\dagger \Phi = C Q_-^\dagger \Phi$, where $C$ is a complex number such that $\vert C \vert < 1$. 
(ii) If either $\mathrm{Im}\, k > 0$ or $f(-\infty) = \pm \infty$, the condition $Q_- \Psi = 0$ in (3.26b) can be replaced by $Q_- \Psi = C Q_+ \Psi$ with $\vert C \vert < 1$ . 

From (3.11), we have
\begin{align*}
&\langle \Phi, U(z,x) Q_+ U(x,z') \Psi \rangle
\\*
& \qquad \qquad \qquad 
= \tau(z,x) \langle \Phi, Q_+ U(z,z') \Psi \rangle 
+ R_l(z,x) \langle \Phi, U(z,x)Q_- U(x,z') \Psi \rangle,
\tag{D1} \\
&\langle \Phi, U(z,x) Q_- U(x,z') \Psi \rangle 
\\*
& \qquad \qquad \qquad 
= \tau(x,z') \langle \Phi, U(z,z') Q_- \Psi \rangle 
+ R_r(x,z') \langle \Phi, U(z,x)Q_+ U(x,z') \Psi \rangle .
\tag{D2}
\end{align*}
Substituting (D1) into the last term of (D2), substituting (D2) into the last term of (D1), and then adding together the resulting two equations, we obtain
\begin{equation}
\psi(x;z,z') = 
\frac{[1 + R_r(x,z')] \tau(z,x)}{1 - R_l(z,x) R_r(x,z')}
\langle \Phi, Q_+ U(z,z') \Psi \rangle
+
\frac{[1 + R_l(z,x)] \tau(x,z')}{1 - R_l(z,x) R_r(x,z')}
\langle \Phi, U(z,z') Q_- \Psi \rangle.
\tag{D3}
\end{equation}
Suppose that 
$Q_- \Psi = C Q_+ \Psi$ with $\vert C \vert < 1$. Then,
\begin{gather}
\langle \Phi, U Q_- \Psi \rangle
= C \langle \Phi, U Q_+ \Psi \rangle 
= C \tau\langle \Phi, Q_+ U \Psi \rangle + C R_l \langle \Phi, U Q_- \Psi \rangle, 
\tag{D4}
\\
\langle \Phi, U(z,z') Q_- \Psi \rangle
= \frac{C \tau(z,z')}{1 - C R_l(z,z')} 
\langle \Phi, Q_+ U(z,z') \Psi \rangle.
\tag{D5}
\end{gather}
Substituting (D5) into (D3) gives
\begin{equation}
\frac{\psi(x;z,z')}{\langle \Phi, Q_+ U(z,z') \Psi \rangle}
=
\frac{[1 + R_r(x,z')] \tau(z,x)}{1 - R_l(z,x) R_r(x,z')}
+
\frac{C \tau(z,z')}{1 - C R_l(z,z')} 
\frac{[1 + R_l(z,x)] \tau(x,z')}{1 - R_l(z,x) R_r(x,z')}.
\tag{D6}
\end{equation}
The first term of (D6) is identical to (B1) with $x_0 = z$. 
Using (B5) and (B7) of Appendix~B, we can see that the second term of (D6) vanishes in the limit $z' \to -\infty$ if $\mathrm{Im}\,k>0$ or $f(-\infty) = \pm \infty$. 
This means that (3.26b) holds with $Q_- \Psi = C Q_+ \Psi$ instead of $Q_- \Psi = 0$.
Note that the condition $\vert C \vert < 1$ can be relaxed to $\vert C \vert \leq 1$, $C \neq -1$ if $f(-\infty) = +\infty$, and $\vert C \vert \leq 1$, $C \neq 1$ if $f(-\infty) = -\infty$. 
In the same way, it can be shown that (3.26a) holds with $Q_+^\dagger \Phi = C Q_-^\dagger \Phi$ instead of $Q_+^\dagger = 0$ if $\mathrm{Im}\, k > 0$ or $f(+\infty) = \pm \infty$.

\section{LIE SUPERALGEBRA $\boldsymbol{osp(1/2)}$}
The orthosymplectic superalgebra\cite{freund} $osp(1/2)$ is defined by the commutation relations
\begin{alignat}{5}
&[J_3, J_\pm] = \pm J_\pm, &\qquad &[J_+, J_-] = 2 J_3, 
\nonumber \\
&[J_3, Q_\pm] = \pm \frac{1}{2}Q_\pm, &\qquad 
&[J_\pm, Q_\pm] = 0, &\qquad &[J_\pm, Q_\mp] = Q_\pm, 
\nonumber \\
&\{Q_+, Q_+\} = -2 J_+, &\qquad &\{Q_-, Q_-\} = 2 J_-, &\qquad
&\{Q_+, Q_-\} = 2 J_3, 
\tag{E1}
\end{alignat}
which also involve the anticommutator defined by $\{A,B\} \equiv A B + B A$. 
The operators $J_\pm$, $J_3$, and $Q_\pm$ of (3.7) can be identified with the elements of $osp(1/2)$.
We can use (4.14) with various representations of $osp(1/2)$. 
A particularly useful representation is given by
\begin{equation}
J_+ = - \frac{1}{2}\, a^\dagger a^\dagger, 
\qquad \! \!
J_- = \frac{1}{2}\, a a, 
\qquad \! \!
J_3 = \frac{1}{2}\, a^\dagger a + \frac{1}{4},
\qquad \! \!
Q_+ = \frac{1}{\sqrt{2}}\, a^\dagger, 
\qquad \! \!
Q_- = \frac{1}{\sqrt{2}}\, a,
\tag{E2}
\end{equation}
where $a$ and $a^\dagger$ are boson annihilation-creation operators satisfying 
$[a, a^\dagger] = 1$.
Expressions of the Green function using the boson representation and other representations of $osp(1/2)$ were studied in Ref.~\onlinecite{algebraic1}. 

The superalgebra $osp(1/2)$ contains a minimal number of elements to fulfill the commutation relations (3.7). The structure of the Schr\"odinger equation described by $osp(1/2)$ is, in some sense, a reduced form of the $SL(3,\mathbf{C})$ structure studied in this paper. 
The general formulas derived in Sec.~VII of this paper cannot be constructed with $osp(1/2)$.

\section{PROOF OF (7.6) and (7.8)}
Equations (7.3) and (7.4) give
\begin{align}
&J_+ \Psi_{p,q} = - (p + q) \Psi_{p+1,q}, \qquad  
K_+ \Psi_{p,q} = p \Psi_{p-1, q+1}, \qquad  \ 
L_+ \Psi_{p,q} = - q \Psi_{p,q-1}, 
\nonumber \\
&J_- \Psi_{p+1,q} = (p + 1) \Psi_{p,q}, \quad \   
K_- \Psi_{p-1, q+1} = (q + 1 ) \Psi_{p,q}, \quad \   
L_- \Psi_{p, q-1} = (p + q - 1) \Psi_{p,q}.
\tag{F1}
\end{align}
From (F1) and (7.5), we can see that the nonzero matrix elements of $J_+$, $K_+$, and $L_+$ are
\begin{align}
&\langle \Psi_{p+1,q}, J_+ \Psi_{p,q} \rangle
= - (p + 1)! \,q!/(p + q - 1)!
= - \langle J_- \Psi_{p+1,q}, \Psi_{p,q} \rangle,
\nonumber \\
&\langle \Psi_{p-1, q+1}, K_+ \Psi_{p,q} \rangle
= p!\,(q + 1)!/(p + q - 1)!
= \langle K_- \Psi_{p-1, q+1}, \Psi_{p,q} \rangle,
\nonumber \\
&\langle \Psi_{p,q-1}, L_+ \Psi_{p,q} \rangle
= - p!\,q! / (p + q - 2)!
= - \langle L_- \Psi_{p,q-1}, \Psi_{p,q} \rangle.
\tag{F2}
\end{align}
It follows from (F2) that $J_\pm^\dagger = - J_\mp$, $L_\pm^\dagger = - L_\mp$, and $K_\pm^\dagger = K_\mp$.
Since the $\Psi_{i,j}$ are eigenvectors of $J_3$, $K_3$, and $L_3$, it is obvious that $J_3^\dagger = J_3$, $K_3^\dagger = K_3$, and $L_3^\dagger = L_3$. 

To prove (7.8), it is sufficient to show that $\langle \Psi_{p',q'}, \Psi_{p,q} \rangle$ given by (7.8) is identical to (7.5). 
We set $\xi \equiv r e^{i \theta}$ and $\mu \equiv s e^{i \phi}$, and rewrite (7.8) using $d\xi_1 d \xi_2 = r dr d\theta$ and $d\mu_2 d\mu_2 = s ds d\phi$.
When we substitute  $\Psi = \Psi_{p,q}$ and $\Phi = \Psi_{p',q'}$ into this expression, the integrals over $\theta$ and $\phi$ produce $\int_0^{2 \pi} e^{i(p-p') \theta} d\theta =2 \pi \delta_{p p'}$ and $\int_0^{2 \pi} e^{i(q-q') \phi} d \phi = 2 \pi \delta_{q q'}$,  and so
\begin{align}
\langle \Psi_{p',q'}, \Psi_{p,q} \rangle
&=
4 \delta_{p p'} \delta_{q q'}\int_0^1 dr \int_0^1 ds 
\ r s \,\delta(1 - r^2 - s^2)
 r^{p'} s^{q'} (1 + p + q) (p+q) r^p s^q
\notag \\*
&=
2\delta_{p p'} \delta_{q q'} (1 + p + q) (p+q)\int_0^1  r^{2p + 1} (1 - r^2)^q \,dr.
\tag{F3}
\end{align}
Setting $t \equiv r^2$ and using the formula
$\int_0^1 t^{x-1} (1 - t)^{y-1} \,dt = \Gamma(x) \Gamma(y)/\Gamma(x+y)$, we find that (F3) coincides with the right-hand side of (7.5).

\section{PROOF OF (7.21)}
Since $\hat{T}(z,z';\xi,\mu) = U(z,z') \mu$ and $U(z,z')= U(3) U(2) U(1)$, we have
\begin{align*}
\hat{T}(z,z';\xi,\mu)
&= U(3) U(2) U(1) \mu
= U(3) U(2) \hat{T}(1; \xi,\mu)
= U(3) \frac{\hat{T}(2;\xi,\mu) \tau(1)}
{1 - \hat{L}(2;\xi) R_r(1)}
\\*
&=
\frac{
\hat{T}(3;\xi,\mu) \tau(2) \tau(1)
}{
[1 - \hat{L}(3;\xi) R_r(2)] [1 - R_l(2) R_r(1)] - \hat{L}(3;\xi)[\tau(2)]^2 R_r(1)
}.
\tag{G1}
\end{align*}
Setting $\xi=0$ and $\mu=1$, we obtain (7.21).

\section{PROOF OF (8.16)}

Assume that $\Psi_0$ is a nonzero vector such that $L_-^{-n} \Psi_0$ exists and $K_+ \Psi_0 = 0$. Let $\Psi_a \equiv L_-^{-n} \Psi_0$. Then, $K_+ \Psi_a = K_+ L_-^{-n} \Psi_0 = L_-^{-n} K_+ \Psi_0 = 0$ and $L_-^n \Psi_a = \Psi_0 \neq 0$. 
As $\Psi_a$ satisfies both $K_+ \Psi_a = 0$ and $L_-^{n} \Psi_a \neq 0$, 
from (8.2) we have $(L_- + K_+)^n U \Psi_a = [(1 + R_r)/\tau]^n U L_-^n \Psi_a$.
Applying $(L_- + K_+)^{-n}$ to both sides of this equation and substituting $\Psi_a = L_-^{-n} \Psi_0$ gives
\begin{equation}
(L_- + K_+)^{-n} U \Psi_0 = 
\left(\frac{\tau}{1 + R_r}\right)^n U L_-^{-n} \Psi_0.
\tag{H1}
\end{equation}
In the same way, it can be shown that
\begin{equation}
\langle \Phi_0, U (L_+ - K_-)^{-n} \cdots \, \rangle 
= 
\left(\frac{\tau}{1 + R_l}\right)^n
\langle \Phi_0, L_+^{-n} U \, \cdots \, \rangle,  
\tag{H2}
\end{equation}
where $\Phi_0$ is a nonzero vector such that $(L_+^{-n})^\dagger \Phi_0$ exists and $K_-^\dagger \Phi_0 = 0$.
Let $\Phi_a \equiv (L_+^{-n})^\dagger \Phi_0$. 
Since $K_-^\dagger \Phi_a = 0$, from (8.4) we have 
$\langle \Phi_a, L_+^n U \cdots \, \rangle
= \tau^n \langle \Phi_a, U L_+^n \cdots \, \rangle$, 
and hence
$\langle \Phi_a, L_+^n U L_+^{-n} \cdots \, \rangle
= \tau^n \langle \Phi_a, U \cdots \, \rangle$.
Substituting $\Phi_a = (L_+^{-n})^\dagger \Phi_0$, we obtain
\begin{equation}
\langle\Phi_0, L_+^{-n} U \cdots \,\rangle
= \frac{1}{\tau^n} 
\langle\Phi_0, U L_+^{-n} \cdots \,\rangle.
\tag{H3}
\end{equation}
Using (H1) -- (H3), we have
\begin{multline}
\langle \Phi_0, U(z,x) (L_+ - K_-)^{-n} U(x,y) (L_- + K_+) ^{-n} U(y,z') \Psi_0 \rangle
\\*
=
\left[
\frac{\tau(z,x)}{1 + R_l(z,x)}
\right]^n
\left[
\frac{\tau(y,z')}{1 + R_r(y,z')}
\right]^n
\langle \Phi_0, L_+^{-n} U(z,z') L_-^{-n} \Psi_0 \rangle
\\*
=
\left[
\frac{\tau(z,x)}{1 + R_l(z,x)}
\frac{\tau(y,z')}{1 + R_r(y,z')}
\frac{1}{\tau(z,z')}
\right]^n
\langle \Phi_0, U(z,z') L_+^{-n} L_-^{-n} \Psi_0 \rangle.
\tag{H4}
\end{multline}
From (8.6) with (H4), it can be seen that (8.16) holds.

\section{DERIVATION OF (8.18) and (8.20)}

The operators $L_-$ and $L_- + K_+$ act on functions of the form $\mu^m h(\xi)$ ($m > 0$) as
\begin{subequations}
\begin{align}
L_- \mu^{m} h(\xi) 
&= \mu^{m + 1} \xi^{1-m} \frac{\partial}{\partial \xi} [\xi^{m} h(\xi)],
\tag{I1}
\\
(L_- + K_+) \mu^{m} h(\xi) 
&= \mu^{m + 1} (1 + \xi)^{1-m}
 \frac{\partial}{\partial \xi} [(1 + \xi)^{m} h(\xi)].
\tag{I2}
\end{align}
\end{subequations}
The right-hand side of (I2) vanishes if $h(\xi) = (1+\xi)^{-m}$, so $L_- + K_+$ is not invertible if $\mu^m (1 + \xi)^{-m}$ is included in its domain. 
We can exclude such functions by restricting the domain of $L_- + K_+$ to functions $\Psi(\xi,\mu)$ satisfying $\vert \Psi(-1,\mu) \vert <\infty$, so we assume $\vert h(-1) \vert < \infty$ in (I2). 
Defining $g(\xi)$ by $(L_- + K_+) \mu^{m} h(\xi) = \mu^{m+1} g(\xi)$, we can invert (I2) as
\begin{equation}
\mu^m h(\xi) = 
(L_- + K_+)^{-1} \mu^{m+1} g(\xi) 
= \mu^m \frac{1}{(1 + \xi)^m}
\int_{-1}^\xi (1 + \xi)^{m-1} g(\xi)\,d\xi.
\tag{I3}
\end{equation}
The lower limit of the integral in (I3) is $-1$ on account of the assumption $\vert h(-1) \vert < \infty$. 
By replacing $m$ with $m-1$, we obtain (8.18b). 
Equation (8.18a) can be derived in an analogous way from (I1). 
The lower limit of the integral in (8.18a) is zero because $\vert h(0) \vert < \infty$.

From (8.18b), we have
\begin{equation}
(L_- + K_+)^{-1}
\left(
\frac{\mu}{1 - c \xi}
\right)^m
=
\left(\frac{\mu}{1 + \xi} \right)^{m-1}
 \int_{-1}^\xi \frac{(1 + \xi)^{m-2}}{(1 - c \xi)^m}\,d\xi
\tag{I4}
\end{equation}
for any complex number $c$.
The integral on the right-hand side of (I4) can be calculated by setting $t = 1 + \xi$ and $\alpha = c/(1 + c)$, and using 
\begin{equation}
\frac{d}{dt}
\left(
\frac{t}{1 - \alpha t}
\right)^{m-1}
= (m - 1) \frac{t^{m - 2}}{(1 - \alpha t)^m}.
\tag{I5}
\end{equation}
As a result, we have\begin{equation}
(L_- + K_+)^{-1}
\left(
\frac{\mu}{1 - c \xi}
\right)^m
=
\frac{1}{m - 1}
\,\frac{1}{1 + c}
\left(
\frac{\mu}{1 - c \xi}
\right)^{m-1}.
\tag{I6}
\end{equation}
Equation (8.20) is obtained from (I6) by iteration.

\section{EXPRESSION OF MULTIPLE SCATTERING IN TERMS OF $\boldsymbol{f}$}

We can formally express the Green function as a perturbative series in terms of $V_\mathrm{S}$,
\begin{align*}
G(x,y) &= G_0(x,y) + \int_{-\infty}^\infty dz_1 \,G_0(x,z_1) V_\mathrm{S}(z_1) G_0(z_1,y) 
\\
& \qquad
+ \int_{-\infty}^\infty  d z_1 \int_{-\infty}^\infty d z_2 \,
G_0(x,z_2) V_\mathrm{S}(z_2) G_0(z_2,z_1)
V_\mathrm{S}(z_1) G_0(z_1,y)
+ \cdots,
\tag{J1}
\end{align*}
where $G_0$ denotes the unperturbed Green function (that is, the Green function for $V_\mathrm{S}=0$). The explicit form of $G_0$ is
\begin{equation}
G_0(x,y) = \frac{1}{2 i k} e^{i k \vert x - y \vert },
\tag{J2}
\end{equation}
as can be checked by substituting into (4.1) with $V_\mathrm{S}=0$. 
Substituting (1.4) and (J2) into the second term on the right-hand side of (J1), and integrating by parts, we have
\begin{multline}
\int_{-\infty}^\infty dz_1\,
G_0(x,z_1) V_\mathrm{S}(z_1) G_0(z_1,y) =
\int_{-\infty}^\infty dz_1\, G_0(x,z_1) [f(z_1)]^2G_0(z_1,y) \\
 + \frac{1}{2ik} \int_{-\infty}^y dz_1 \,  f(z_1) \, e^{ik(x - 2 z_1 + y )}
  - \frac{1}{2ik} \int_x^\infty dz_1 \, f(z_1) \, e^{- ik(x - 2 z_1 + y)}.
\tag{J3}
\end{multline}
[Since we are treating $V_\mathrm{S}$ as a perturbation, we have assumed $f(\pm \infty)=0$.]
Similarly, from the third term on the right-hand side of (J1), \begin{align*}
&\int_{-\infty}^\infty  d z_1 \int_{-\infty}^\infty d z_2 \,
G_0(x,z_2) V_\mathrm{S}(z_2) G_0(z_2,z_1)
V_\mathrm{S}(z_1) G_0(z_1,y) 
\\
& \qquad
=
-\int_{-\infty}^\infty dz_1\, G_0(x,z_1) [f(z_1)]^2G_0(z_1,y)
-\frac{1}{2 i k}\iint_{A_2} dz_1 dz_2
\, f(z_1) f(z_2) 
\, e^{- ik(x - 2 z_2 + 2 z_1 - y)}
\\
&\qquad \qquad
-\frac{1}{2 i k}\iint_{B_2} dz_1 dz_2
\, f(z_1) f(z_2) 
\, e^{ik(x - 2 z_2 + 2 z_1 - y)} 
+ \mbox{(higher order in $f$)},
\tag{J4}
\end{align*}
where $A_2$ denotes the region $z_1 \leq y$, $x \leq z_2$, and $B_2$ denotes the region $y \leq z_1$, $z_2 \leq z_1$, $z_2 \leq x$. [We are assuming $y \leq x$, so $z_1 \leq z_2$ in $A_2$. We can write $\iint_{A_2} dz_1 dz_2 = \int_{-\infty}^y dz_1 \int_x^\infty dz_2$ and $\iint_{B_2} dz_1 dz_2 = \int_y^\infty dz_1 \int_{-\infty}^{\min(x,z_1)} dz_2 = \int_{-\infty}^x dz_2 \int_{\max(y,z_2)}^\infty dz_1$.]
The first term on the right-hand side of (J4) cancels the first term on the right-hand side of (J3). This kind of cancellation happens at every order in $f$, and integrals involving $[f(z)]^2$ do not remain in (J1). Substituting (J3), (J4), and so on into (J1) and writing
\begin{equation*}
e^{\pm ik(x - 2 z_1 + y )}
= e^{\pm ik(x - z_2)} e^{\pm ik(y - z_1)},
\qquad
e^{\pm ik(x - 2 z_2 + 2 z_1 - y)}
= e^{\pm ik(x - z_2)} e^{\pm ik(z_1 - z_2)} e^{\pm ik(z_1 - y)},
\end{equation*}
we obtain (10.1). 

For general integer $n$, the terms of order $2n$ in (10.1) have the form
\begin{multline}
(-1)^n\int \cdots \int_{A_{2n}} dz_1 \cdots dz_n
\, f(z_1) \cdots f(z_{2n})  
\, e^{ik(x - 2z_{2n} + 2 z_{2n-1} - \cdots - 2 z_2 + 2 z_1 - y)}
\notag \\
 + (-1)^n\int \cdots \int_{B_{2n}} dz_1 \cdots dz_n
\, f(z_1) \cdots f(z_{2n})  
\, e^{- ik(x - 2z_{2n} + 2 z_{2n-1} - \cdots - 2 z_2 + 2 z_1 - y)},
\tag{J5}
\end{multline}
and the terms of order $2n+1$ are
\begin{multline}
(-1)^n\int \cdots \int_{A_{2n+1}} dz_1 \cdots dz_{2n+1}
\, f(z_1) \cdots f(z_{2n+1})  
\, e^{ik(x - 2z_{2n+1} + 2z_{2n} - \cdots + 2 z_2 - 2 z_1 + y)}
\notag \\
+ (-1)^{n+1}\int \cdots \int_{B_{2n+1}} dz_1 \cdots dz_{2n+1}
\, f(z_1) \cdots f(z_{2n+1})  
\, e^{-ik(x - 2z_{2n+1} + 2z_{2n} - \cdots + 2 z_2 - 2 z_1 + y)},
\tag{J6}
\end{multline}
where $A_{2n}$, $B_{2n}$, $A_{2n+1}$, and $B_{2n+1}$ denote the following regions:
\begin{align*}
A_{2n} &: \quad \
y \leq z_1, \quad z_{2i} \leq z_{2i - 1} \ (1 \leq i \leq n), \quad 
z_{2i} \leq z_{2i + 1} \ (1 \leq i \leq n-1), \quad z_{2n} \leq x,
\\
B_{2n} &: \quad \
z_1 \leq y, \quad z_{2i - 1} \leq  z_{2i} \ (1 \leq i \leq n), \quad 
z_{2i + 1} \leq z_{2i} \ (1 \leq i \leq n-1), \quad x \leq z_{2n},
\\
A_{2n+1} &: \quad \
z_1 \leq y, \quad z_{2i - 1} \leq  z_{2i} \ (1 \leq i \leq n), \quad 
z_{2i + 1} \leq z_{2i} \ (1 \leq i \leq n), \quad z_{2n+1} \leq x,
\\
B_{2n+1} &: \quad \
y \leq z_1, \quad z_{2i} \leq z_{2i - 1} \ (1 \leq i \leq n), \quad 
z_{2i} \leq z_{2i + 1} \ (1 \leq i \leq n), \quad x \leq z_{2n+1}.
\end{align*}
Each term of (J5) and (J4) corresponds to a path like the one in Fig.~5.

\end{document}